\documentclass{article}

\usepackage[final]{doodle}

\usepackage[utf8]{inputenc}
\usepackage[T1]{fontenc}
\usepackage{url}
\usepackage{booktabs}
\usepackage{amsfonts}
\usepackage{nicefrac}
\usepackage{microtype}
\usepackage{xcolor}
\usepackage{amssymb}
\usepackage{amsmath}
\usepackage{natbib}

\title{Sliced Tensor Operators for\\Large-Scale Topology Optimization}

\author{
  Pedro Zuidberg Dos Martires \\
  Department of Computer Science \\
  Örebro University \\
  Sweden
  \And
  Gustavo Assis da Silva \\
  Department of Mechanical Engineering \\
  Örebro University \\
  Sweden
}

\usepackage{graphicx}%
\usepackage{multirow}%
\usepackage{amsmath,amssymb,amsfonts}%
\usepackage{amsthm}%
\usepackage{mathrsfs}%
\usepackage[title]{appendix}%
\usepackage{xcolor}%
\usepackage{textcomp}%
\usepackage{manyfoot}%
\usepackage{booktabs}%
\usepackage{algorithm}%
\usepackage{algorithmicx}%
\usepackage{algpseudocode}%
\algrenewcommand{\algorithmiccomment}[1]{\hfill\textcolor{gray}{\(\triangleright\) #1}}

\usepackage{listings}%
\usepackage{placeins}
\usepackage{graphicx}
\usepackage{subcaption}

\usepackage[dvipsnames]{xcolor}
\usepackage[
    colorlinks=true,
    linkcolor=BrickRed,
    citecolor=BrickRed,
    urlcolor=BrickRed
]{hyperref}
\usepackage{xspace}
\usepackage{adjustbox}

\usepackage{nicefrac}

\usepackage{tikz}
\usepackage{xcolor}
\usetikzlibrary{arrows.meta,calc,fit,backgrounds}

\definecolor{topaxInk}{HTML}{27313A}
\definecolor{topaxMuted}{HTML}{66717B}
\definecolor{topaxLine}{HTML}{AEB7BF}
\definecolor{topaxPale}{HTML}{F3F5F6}
\definecolor{topaxBlue}{HTML}{4C78A8}
\definecolor{topaxBluePale}{HTML}{EAF2F8}
\definecolor{topaxGreen}{HTML}{4F806F}
\definecolor{topaxGreenPale}{HTML}{ECF4F0}
\definecolor{topaxOrange}{HTML}{C97818}
\definecolor{topaxOrangePale}{HTML}{FFF0D8}

\usepackage[frozencache=true,cachedir=_minted]{minted}
\usepackage{fvextra} 
\usepackage{float}   
\usepackage{nicefrac}
\usepackage{mathtools}

\DeclareRobustCommand{\topax}{\textsf{topax}}

\newcommand{\cf}{c.f.\xspace}

\newcommand{\elementstiffness}{\ensuremath{J}}
\newcommand{\stiffness}{\ensuremath{K}}

\newcommand{\transpose}{\ensuremath{\mathsf{T}}}

\newcommand{\fvec}{\ensuremath{\boldsymbol{f}}}

\newcommand{\zvec}{\ensuremath{\boldsymbol{z}}}
\newcommand{\pvec}{\ensuremath{\boldsymbol{p}}}
\newcommand{\vvec}{\ensuremath{\boldsymbol{v}}}
\newcommand{\rvec}{\ensuremath{\boldsymbol{r}}}
\newcommand{\rhovec}{\ensuremath{\boldsymbol{\rho}}}
\newcommand{\uvec}{\ensuremath{\boldsymbol{u}}}

\newcommand{\thetavec}{\ensuremath{\boldsymbol{\theta}}}

\begin{document}

\maketitle

\begin{abstract}
    Large-scale three-dimensional topology optimization is constrained by both the cost of repeatedly solving the governing finite-element equations and the memory required by these computations.
    Memory consumption can be controlled through specialized low-level implementations, but developing, optimizing, and extending such code requires substantial effort and hardware-specific expertise.
    High-level tensor programming frameworks, which support automatic differentiation, offer an attractive alternative: numerical operations can be expressed compactly, while their compilers generate highly optimized code for CPUs and GPUs.
    However, a direct tensor formulation creates intermediate arrays that can quickly exhaust the available memory.
    We address this limitation through memory-bounded tensor operators for matrix-free finite-element analysis and Galerkin multigrid.
    The fine-grid stiffness action is evaluated in overlapping spatial slices, and on coarse grids, the Galerkin operator is retained in a factorized representation consisting of fine-grid stiffness interpolation coefficients and precomputed local operators.
    This representation permits its action to be evaluated without materializing a spatially varying stiffness operator on every coarse grid, while preserving the exact Galerkin hierarchy.
    We implement these formulations in \topax{}, a modular JAX library for two- and three-dimensional density-based topology optimization supporting CPUs and GPUs.
    On a single NVIDIA GH200, \topax{} performs 50 design iterations for a problem with more than $809$ million displacement degrees of freedom in approximately $24.5$ minutes.
    Across the common GPU benchmark, it is upt to $13.6$ times faster than the fastest competing implementation and uses less than half the device memory for the largest problem.
\end{abstract}

\section{Introduction}
\label{sec:intro}

The rapid development of machine learning has led to substantial advances in software and hardware infrastructure for high-performance tensor computations.
Modern numerical frameworks, such as JAX~\citep{jax2018github} and PyTorch~\citep{Paszke2019PyTorchAI}, provide access to highly optimized computational kernels, just-in-time compilation, and automatic differentiation.
These capabilities also offer opportunities for accelerating scientific computing, provided that numerical algorithms can be expressed through suitable computational primitives.

Finite-element analysis is one example of a computational method that can benefit from this infrastructure.
Conventional implementations rely on element connectivity, sparse matrix assembly, and sparse linear algebra~\citep{cecka2011assembly}, which often involve irregular memory access and indexing operations.
For Cartesian meshes, however, the regularity of the discretization can be exploited to express local numerical operators through structured tensor operations.
For instance, finite-difference discretizations of differential operators can be implemented using convolutional kernels~\citep{he2019mgnet}, while matrix-free finite-element formulations can exploit regular element connectivity to avoid global stiffness matrix assembly~\citep{martinez2015efficient}.
Such formulations can leverage optimized CPU and GPU implementations without requiring specialized low-level code.

Density-based topology optimization provides a particularly demanding application of finite-element analysis~\citep{sigmund2013topology}.
The material distribution changes throughout the optimization, requiring the solution of large linear elasticity problems at every design iteration.
For high-resolution discretizations, these repeated solves constitute a major computational expense.
Matrix-free methods reduce memory requirements by evaluating the stiffness action without assembling the global sparse matrix.
Combined with multigrid preconditioning, they provide an effective approach for large-scale topology optimization, as demonstrated by previous GPU-accelerated implementations~\citep{dick2011real,wu2015system}.
However, expressing finite-element operators through tensor operations can introduce large intermediate arrays, limiting the problem sizes that can be handled efficiently.
Similar memory challenges arise in other applications of tensor computations, where techniques such as tiling and index slicing have been used to reduce data movement and decompose large operations into smaller computations~\citep{ziogas2022deinsum,huang2021efficient}.

In this work, we address this limitation by expressing the fine-grid stiffness action through convolutions and structured tensor transformations, and evaluating it in overlapping slices.
This avoids explicit element connectivity during operator evaluation while bounding the memory required for intermediate computations.
We extend this approach to the Galerkin coarse-grid operators used in multigrid preconditioning.
Each coarse-element stiffness matrix is decomposed into a weighted combination of precomputed local operators, allowing its action to be evaluated directly through tensor contractions without materializing individual stiffness matrices for each coarse element.
The weights are obtained from the original fine-grid material distribution, while the precomputed operators are shared across elements at each multigrid level.
Processing these contractions in bounded batches further reduces intermediate memory requirements.

We implement these methods in \topax{}, a modular topology-optimization library developed using JAX.
The library combines the proposed operators with automatic differentiation and just-in-time compilation, supports execution on CPUs and GPUs, and allows different optimization algorithms, filtering methods, projection functions, and solver components to be selected independently.

The main contributions of this work are:
\begin{enumerate}
    \item A convolution-based matrix-free finite-element formulation for Cartesian meshes, expressed through regular tensor operations that can leverage optimized CPU and GPU implementations.

    \item A tensor slicing strategy that limits the intermediate memory requirements of the fine-grid stiffness action while preserving the underlying finite-element operator.

    \item An implicit Galerkin formulation that evaluates coarse-grid stiffness actions directly from fine-grid material coefficients and precomputed local operators, together with a batched evaluation strategy for controlling intermediate memory consumption.

    \item \topax{}, a modular JAX implementation integrating these operators into a multigrid preconditioned topology-optimization framework with support for two- and three-dimensional problems.
\end{enumerate}

The proposed formulations and their implementation are evaluated through numerical experiments that assess their impact on computational performance and memory consumption.
The results demonstrate substantial performance improvements over recent state-of-the-art topology optimization implementations, while highlighting the trade-offs between execution time and memory requirements.
In our largest experiment, \topax{} completes 50 design iterations for a problem with more than $809$ million displacement degrees of freedom on a single GPU in approximately $24.5$ minutes, with a peak memory consumption of $83.4\,\mathrm{GiB}$.


\section{Preliminaries}
\label{sec:prelim}

\subsection{Density-based topology optimization}

This work considers the standard compliance minimization problem subject to a volume constraint \citep{bendsoe2004topology}.
A density-based approach is adopted to represent both the design variables and the physical relative densities within the design domain \citep{sigmund2013topology}.
In this formulation, the design domain is discretized using the standard displacement-based finite element method \citep{bathe2014finite}, and each finite element is associated with a design variable $\theta_e \in [0,1]$.
The collection of all design variables is represented by the vector $\thetavec \in \mathbb{R}^{N_e}$, where $N_e$ denotes the total number of elements in the discretization.

The design variables are processed using either a density filter based on a linear convolution operator ~\citep{bourdin2001filters,bruns2001topology} or a Helmholtz-type filter~\citep{lazarov2011filters}, optionally followed by a projection operation~\citep{wang2011projection}.
This process yields the physical relative density vector $\rhovec\in\mathbb{R}^{N_e}$, whose components $\rho_e$ represent element-wise constant relative densities used in the finite-element analysis, constraint evaluation, and sensitivity computations.
Since the physical densities are obtained by filtering and, optionally, projecting the design variables, $\rhovec$ may be regarded as a function of $\thetavec$.
Alternatively, sensitivity filtering may be used \citep{sigmund200199}. In this formulation, the physical densities are identified directly with the design variables, $\rhovec=\thetavec$, and the objective and constraint sensitivities are filtered before being passed to the optimization algorithm.

Considering linear elastic behavior under static loading conditions, the compliance minimization problem may be stated as
\begin{equation} 
\begin{aligned} 
\min_{\thetavec} \quad & c(\thetavec) = \fvec^{\transpose}\uvec(\thetavec) \\ 
\text{s.t.} \quad & \frac{1}{N_e}\sum_{e=0}^{N_e-1}\rho_e (\thetavec) \leq V_f \\ 
& \stiffness(\thetavec)\uvec(\thetavec) = \fvec \\
& 0 \le \theta_e \le 1, \qquad e\in\{0,\ldots,N_e-1\} 
\end{aligned},
\label{eq:topopt} 
\end{equation}
where $c$ is the structural compliance, $\uvec$ is the global displacement vector, $\fvec$ is the external load vector, $\stiffness$ is the global stiffness matrix, and $V_f$ is the prescribed volume fraction.
The dependence of these quantities on the design variables is explicit in Equation~\ref{eq:topopt}.
In the remainder of this work, such dependencies are omitted whenever no ambiguity arises.

The element stiffness matrices are interpolated using the \textit{solid isotropic material with penalization} (SIMP) approach~\citep{bendsoe1999material,sigmund2007morphology}, according to
\begin{equation}
    \stiffness_e
    =
    \underbrace{
        \left( \rho_{\min} + \rho_e^{\,p} \left(1-\rho_{\min}\right) \right)
    }_{\eqcolon w_e}
    \elementstiffness_{e},
    \label{eq:simp}
\end{equation}
where $\elementstiffness_{e}$ is the stiffness matrix of element $e$ for the solid material, $p$ is the penalization exponent, and $\rho_{\min}$ is a small positive parameter introduced to prevent singular stiffness matrices.
The global stiffness matrix is defined through the standard finite-element assembly~\citep{bathe2014finite}
\begin{equation}
    \stiffness = \sum_{e=0}^{N_e-1} L_e^{\transpose} \stiffness_e L_e, \label{eq:global_stiffness_assembly}
\end{equation}
where $L_e$ is the Boolean localization matrix that relates the global and element degree-of-freedom arrangements.
In particular, the element displacement vector $\uvec_e$ is obtained from the global displacement vector as $\uvec_e=L_e\uvec$.
Equation~\ref{eq:global_stiffness_assembly} defines the global stiffness matrix algebraically and does not imply its explicit assembly, which is avoided by the matrix-free formulations introduced in the following sections.

\subsection{Multigrid preconditioned conjugate gradient solver}\label{sec:mgcg}

For large-scale topology optimization problems, solving the finite-element equilibrium problem in Equation~\ref{eq:topopt} using direct methods becomes computationally expensive due to the memory requirements and computational cost associated with matrix factorization.
Iterative methods, such as the conjugate gradient (CG) method, provide an alternative for solving the resulting large, sparse, symmetric positive-definite linear systems.
However, the convergence of CG depends on the condition number of the stiffness matrix, which may be particularly unfavorable in density-based topology optimization due to the large variations in material stiffness throughout the design domain.

Preconditioning is therefore employed to improve the convergence behavior of CG by reducing the effective condition number of the linear system.
In particular, geometric multigrid methods are well suited to large-scale finite-element problems, as they combine smoothing operations that reduce high-frequency error components with coarse-grid corrections that address low-frequency errors~\citep{trottenberg2001multigrid}.
In this paper, we use a geometric multigrid V-cycle as a preconditioner within the CG iteration, resulting in the multigrid preconditioned conjugate gradient (MGPCG) method~\citep{amir2014multigrid}, summarized in Algorithm~\ref{alg:mgcg}.

\begin{algorithm}[t]
\caption{Multigrid preconditioned conjugate gradient}
\label{alg:mgcg}
\begin{algorithmic}[1]
\Function{MGPCG}{$\stiffness,\fvec,\uvec_0$}
    \State $\rvec_0 \gets \fvec-\stiffness\uvec_0$
    \State $\zvec_0 \gets \mathcal{G}(\rvec_0,0)$
    \State $\pvec_0 \gets \zvec_0$
    \State $j \gets 0$
    \While{stopping criterion not met}
        \State $\vvec_j \gets \stiffness\pvec_j$
        \State $\alpha_j \gets
        \nicefrac{\rvec_j^{\transpose}\zvec_j} 
        {\pvec_j^{\transpose}\vvec_j}$\label{alg:mgcg:line:precon_residual_num}
        \State $\uvec_{j+1} \gets \uvec_j+\alpha_j\pvec_j$
        \State $\rvec_{j+1} \gets \rvec_j-\alpha_j\vvec_j$
        \State $\zvec_{j+1} \gets \mathcal{G}(\rvec_{j+1},0)$
        \State $\beta_j \gets
        \nicefrac{\rvec_{j+1}^{\transpose}\zvec_{j+1}}
        {\rvec_j^{\transpose}\zvec_j}$ \label{alg:mgcg:line:precon_residual_denom}
        \State $\pvec_{j+1} \gets
        \zvec_{j+1}+\beta_j\pvec_j$
        \State $j \gets j+1$
    \EndWhile
    \State \Return $\uvec_j$
\EndFunction
\end{algorithmic}
\end{algorithm}

In Algorithm~\ref{alg:mgcg}, $\rvec_j$, $\zvec_j$, and $\pvec_j$ denote the residual, preconditioned residual, and search direction at iteration $j$, respectively.
The preconditioning operation $\mathcal{G}(\rvec_j,0)$ is evaluated using the recursive geometric multigrid V-cycle defined in Algorithm~\ref{alg:vcycle}.
Convergence is typically assessed using the relative residual norm.
The loop is exited when $\|\rvec_j\|_2 / \|\fvec\|_2 \leq \varepsilon$, where $\varepsilon$ denotes a prescribed tolerance.
In practice, a maximum number of loop iterations is additionally imposed to prevent excessive computational effort in cases where the desired tolerance is not reached.

\begin{algorithm}[t]
\caption{Recursive geometric multigrid V-cycle}
\label{alg:vcycle}
\begin{algorithmic}[1]
\Function{$\mathcal{G}$}{$\rvec^{(\ell)},\ell$}
    \If{$\ell=L-1$}
        \State $\zvec^{(\ell)} \gets  \left( \stiffness^{(\ell)}\right)^{-1} \rvec^{(\ell)}$
        \Comment{Direct solve}
    \Else
        \State $\zvec^{(\ell)} \gets \boldsymbol{0}$
        \State $\zvec^{(\ell)} \gets
        \mathcal{S}^{(\ell)}(\rvec^{(\ell)},\zvec^{(\ell)})$
        \Comment{Pre-smoothing}
        \State $\tilde{\rvec}^{(\ell)}
        \gets \rvec^{(\ell)}-\stiffness^{(\ell)}\zvec^{(\ell)}$
        \State $\rvec^{(\ell+1)}
        \gets (P^{(\ell)})^{\transpose}\tilde{\rvec}^{(\ell)}$
        \Comment{Restriction}
        \State $\zvec^{(\ell+1)}
        \gets \mathcal{G}(\rvec^{(\ell+1)},\ell+1)$
        \Comment{Recursive call}
        \State $\zvec^{(\ell)}
        \gets \zvec^{(\ell)}+P^{(\ell)}\zvec^{(\ell+1)}$
        \Comment{Prolongation}
        \State $\zvec^{(\ell)} \gets
        \mathcal{S}^{(\ell)}(\rvec^{(\ell)},\zvec^{(\ell)})$
        \Comment{Post-smoothing}
    \EndIf
    \State \Return $\zvec^{(\ell)}$
\EndFunction
\end{algorithmic}
\end{algorithm}

In Algorithm~\ref{alg:vcycle}, $\ell$ denotes the current multigrid level, with $\ell=0$ corresponding to the finest level and $\ell=L-1$ to the coarsest level, where $L$ is the total number of levels in the V-cycle.
The vectors $\rvec^{(\ell)}$ and $\zvec^{(\ell)}$ denote the right-hand side and correction at level $\ell$, respectively, while $\stiffness^{(\ell)}$ is the corresponding stiffness matrix.
The operator $\mathcal{S}^{(\ell)}$ denotes the smoothing procedure, applied before and after the coarse-grid correction.
The operator $P^{(\ell)}$ prolongates vectors from level $\ell{+}1$ to level $\ell$, while its transpose $\left(P^{(\ell)}\right)^{\transpose}$ is used to restrict vectors from level $\ell$ to level $\ell+1$.
Common smoothing schemes include Gauss--Seidel and damped Jacobi iterations~\citep{trottenberg2001multigrid}.
The latter is particularly well suited to GPU implementations, as its updates can be evaluated independently and efficiently in parallel.
The linear system at level $L{-}1$ is solved using a sparse direct solver, typically through sparse Cholesky factorization.
The level-dependent stiffness matrices are defined recursively using the Galerkin coarse-grid construction~\citep{trottenberg2001multigrid},
\begin{equation}
\stiffness^{(\ell+1)}
=
\left(P^{(\ell)}\right)^{\transpose}
\stiffness^{(\ell)}
P^{(\ell)},\qquad \ell\in\{0,\ldots,L-2\},
\label{eq:galerkin}
\end{equation}
starting from the fine-grid stiffness matrix $\stiffness^{(0)}$.
A structured dyadic multigrid hierarchy is employed, in which the mesh is coarsened by a factor of two along each spatial direction at every level.
Consequently, each element at level $\ell{+}1$ contains four elements from level $\ell$ in two dimensions and eight elements in three dimensions.


\section{Slicing matrix-free linear operators}
\label{sec:free_slicing}

The repeated solution of large-scale finite-element systems constitutes a major computational expense in density-based topology optimization.
Although the MGPCG method avoids the factorization of the fine-grid stiffness matrix, each iteration requires repeated evaluations of the matrix--vector product $\stiffness\uvec$.
For large problems, explicitly assembling and storing the global stiffness matrix $\stiffness$ can impose substantial memory requirements.
Matrix-free finite-element methods avoid this storage by evaluating the stiffness action directly from element-level contributions~\citep{kronbichler2012generic}.

Efficient GPU execution further benefits from regular computational structures that expose parallelism and avoid irregular memory access.
We consider uniform Cartesian meshes discretized using bilinear quadrilateral elements in two dimensions and trilinear hexahedral elements in three dimensions.
The regular connectivity allows the finite-element matrix--vector product to be expressed through convolution-like tensor operations, eliminating the need for explicit element-to-node and node-to-element connectivity arrays during operator application.
For the sake of clarity, our formulation of the linear operation $\stiffness\uvec$ is presented in detail for the two-dimensional case.
Its extension to three dimensions is then straightforward.

Consider a structured Cartesian mesh comprising $N_x\times N_y$ bilinear quadrilateral elements.
The element-wise stiffness interpolation coefficients appearing in Equation~\ref{eq:simp} are collected in the tensor $W\in\mathbb{R}^{N_x\times N_y}$, with $W_{ij}=\rho_{\min}+\rho_{ij}^{\,p}(1-\rho_{\min})$.
The subscripts $(i,j)$ denote the element index (\cf subscript $e$ in Equation~\ref{eq:simp}) in the structured mesh, with $i\in \{0,\ldots,N_x{-}1\}$ and $j\in \{0,\ldots,N_y{-}1\}$.

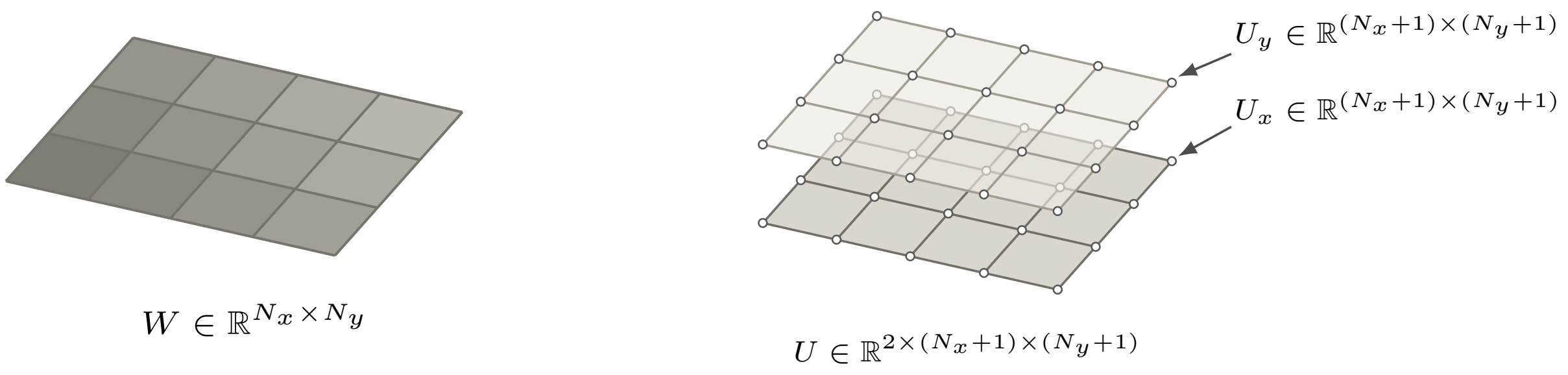
\begin{figure}[t]
    \centering

    \begin{subfigure}[c]{0.3\textwidth}
        \centering
        \resizebox{\linewidth}{!}{%
            \begin{tikzpicture}[line join=round, font=\small]
  \definecolor{elementBase}{HTML}{E8E5DF}
  \definecolor{elementLine}{HTML}{77736D}

  \begin{scope}[x={(.62cm,-.14cm)},y={(.32cm,.36cm)}]

    \foreach \i in {0,...,3}{
      \foreach \j in {0,...,2}{
        \pgfmathtruncatemacro{\shadelevel}{8+10*(\i+\j)}
        \fill[elementBase!\shadelevel!elementLine]
          (\i,\j)--(\i+1,\j)--(\i+1,\j+1)--(\i,\j+1)--cycle;
      }
    }

    \foreach \i in {0,...,4}{
      \draw[elementLine,line width=.55pt] (\i,0)--(\i,3);
    }
    \foreach \j in {0,...,3}{
      \draw[elementLine,line width=.55pt] (0,\j)--(4,\j);
    }

  \end{scope}

\node[font=\scriptsize,anchor=north]
    at (1.88,-0.8) {$W\in\mathbb R^{N_x\times N_y}$};

\end{tikzpicture}%
        }
    \end{subfigure}%
    \hfill
    \begin{subfigure}[c]{0.52\textwidth}
        \centering
        \resizebox{\linewidth}{!}{%
            \begin{tikzpicture}[
  font=\small,line join=round,>=Latex,
  nodepoint/.style={circle,draw=black!65,fill=white,inner sep=0pt,minimum size=2.0pt}
]
  \definecolor{lowerLayer}{HTML}{D4D0C9}
  \definecolor{upperLayer}{HTML}{EEEAE3}
  \definecolor{lowerGrid}{HTML}{736E66}
  \definecolor{upperGrid}{HTML}{A49D92}

  \begin{scope}[x={(.62cm,-.14cm)},y={(.32cm,.36cm)}]
    \fill[lowerLayer,opacity=.88]
      (0,0)--(4,0)--(4,3)--(0,3)--cycle;

    \foreach \i in {0,...,4}{
      \draw[lowerGrid,line width=.55pt] (\i,0)--(\i,3);
    }
    \foreach \j in {0,...,3}{
      \draw[lowerGrid,line width=.55pt] (0,\j)--(4,\j);
    }
    \foreach \i in {0,...,4}{
      \foreach \j in {0,...,3}{
        \node[nodepoint] at (\i,\j) {};
      }
    }
  \end{scope}

  \begin{scope}[
    shift={(0,.66)},
    x={(.62cm,-.14cm)},
    y={(.32cm,.36cm)}
  ]
    \fill[upperLayer,opacity=.66]
      (0,0)--(4,0)--(4,3)--(0,3)--cycle;

    \foreach \i in {0,...,4}{
      \draw[upperGrid,line width=.55pt] (\i,0)--(\i,3);
    }
    \foreach \j in {0,...,3}{
      \draw[upperGrid,line width=.55pt] (0,\j)--(4,\j);
    }
    \foreach \i in {0,...,4}{
      \foreach \j in {0,...,3}{
        \node[nodepoint] at (\i,\j) {};
      }
    }
  \end{scope}

  \node[font=\scriptsize,anchor=west,inner sep=.5pt]
    (UyLabel) at (3.94,1.60)
    {$U_y\in\mathbb R^{(N_x+1)\times(N_y+1)}$};

  \draw[-{Latex[length=1.6mm]},black!70,line width=.55pt]
    (UyLabel.south west) -- (3.49,1.23);

  \node[font=\scriptsize,anchor=west,inner sep=.5pt]
    (UxLabel) at (3.94,.97)
    {$U_x\in\mathbb R^{(N_x+1)\times(N_y+1)}$};

  \draw[-{Latex[length=1.6mm]},black!70,line width=.55pt]
    (UxLabel.south west) -- (3.49,.56);

\node[font=\scriptsize,anchor=north]
    at (1.72,-0.8)
    {$U\in\mathbb R^{2\times(N_x+1)\times(N_y+1)}$};

\end{tikzpicture}%
        }
    \end{subfigure}

    \caption{Element-wise and nodal tensor representations on a structured quadrilateral mesh.
    Left: element-wise stiffness interpolation coefficients $W\in\mathbb{R}^{N_x\times N_y}$.
    Right: two-component nodal displacement field $U\in\mathbb{R}^{2\times(N_x+1)\times(N_y+1)}$.}

    \label{fig:matrixfree_fields}
\end{figure}

For the tensor-based formulation, the nodal displacement field is conveniently represented by $U\in\mathbb{R}^{2\times(N_x+1)\times(N_y+1)}$.
The first dimension distinguishes the two displacement components, while the remaining dimensions correspond to the nodal indices of the structured mesh.
Accordingly, $U$ comprises the displacement fields $U_x,U_y\in\mathbb{R}^{(N_x+1)\times(N_y+1)}$, where $U_x$ and $U_y$ collect the nodal displacement components in the $x$- and $y$-directions, respectively.
The global displacement vector $\uvec$ and the tensor $U$ represent the same nodal displacement field, with the displacement components arranged according to different indexing conventions.
Figure~\ref{fig:matrixfree_fields} illustrates these representations, highlighting the distinction between the element-wise stiffness coefficients and the two-component nodal displacement field.

\subsection{Tensor-based matrix-free operator}

Consider an arbitrary quadrilateral element $e$ indexed by $(i,j)$ in the structured mesh.
We denote its stiffness interpolation coefficient $W_{ij}$ by $w_e$ for convenience.
Each element has four nodes, with two displacement degrees of freedom per node.
We collect these eight degrees of freedom in the element displacement vector $\uvec_e\in\mathbb{R}^{8}$.
Throughout the formulation, we adopt a component-first ordering of the local degrees of freedom: the four $x$-displacement components precede the four $y$-displacement components, with the nodes ordered first along the $x$-direction and then along the $y$-direction.

Because the mesh is uniform and all elements share the same geometry as well as solid-material properties, the matrices $\elementstiffness_{e} \in \mathbb{R}^{8\times8}$ in Equation~\ref{eq:simp} are identical for all $e$.
We therefore drop the subscript $e$ and write
\begin{align}
    \stiffness_e &= w_e\elementstiffness.
    \label{eq:homogeneous_Ke}
\end{align}
The element stiffness matrix maps the element displacement vector to the element force vector according to
\begin{align}
    \fvec_e &= \stiffness_e\uvec_e = w_e\elementstiffness \uvec_e, \label{eq:element_stiffness_action}
\end{align}
where $\fvec_e\in\mathbb{R}^{8}$ collects the eight force contributions associated with the element degrees of freedom.
While $\elementstiffness$ is common to all elements, the stiffness interpolation coefficient $w_e$ and the element displacement vector $\uvec_e$ vary across the mesh.

Consider now the unweighted element force contributions,
\begin{align}
    \mathbf{h}_e &= \elementstiffness\uvec_e.\label{eq:matrixfree_unweighted_element}
\end{align}
Collecting these contributions over all elements gives the tensor $H\in\mathbb{R}^{8\times N_x\times N_y}$, where $H_{aij}$ denotes the $a$th unweighted force contribution of the element indexed by $(i,j)$.

\begin{figure}[h]
    \centering

    \begin{subfigure}[c]{0.39\textwidth}
        \centering
        \resizebox{0.9\linewidth}{!}{%
            \definecolor{oiOrange}{HTML}{E69F00}
\definecolor{oiSkyBlue}{HTML}{56B4E9}
\definecolor{oiGreen}{HTML}{009E73}
\definecolor{oiYellow}{HTML}{F0E442}
\definecolor{oiBlue}{HTML}{0072B2}
\definecolor{oiVermillion}{HTML}{D55E00}
\definecolor{oiPurple}{HTML}{CC79A7}
\definecolor{oiBlack}{HTML}{000000}
\begin{tikzpicture}[
  x=1cm,y=1cm,>=Latex,
  font=\small,
  edge/.style={draw=black!58,line width=.55pt},
  arrow/.style={-Latex,draw=black!65,line width=.8pt},
  label/.style={font=\footnotesize},
]
\begin{scope}[shift={(0,1.55)}]
  \foreach \n in {0,...,7}{
    \draw[edge] (.217142857*\n,0)--(.217142857*\n,-1.52);
    \draw[edge] (0,-.217142857*\n)--(1.52,-.217142857*\n);
  }
  \foreach \i in {0,...,7}{
    \foreach \j in {0,...,7}{
      \filldraw[draw=black!75,fill=black!80,line width=.25pt]
        (.217142857*\i,-.217142857*\j) circle[radius=.025];
    }
  }
  \node[label] at (.76,-1.85) {$\elementstiffness \in\mathbb R^{8\times 8}$};
  \draw[arrow] (1.77,-.75)--(2.50,-.75);
  \node[label,above] at (2.135,-.70) {reshape};
\end{scope}
\newcommand{\kernelcube}[4]{%
  \begin{scope}[shift={(#1,#2)}]
    \fill[#4,opacity=.11] (.21,.18)--(.83,.18)--(.83,.80)--(.21,.80)--cycle;
    \fill[#4,opacity=.16] (0,0)--(.62,0)--(.83,.18)--(.21,.18)--cycle;
    \fill[#4,opacity=.20] (.62,0)--(.83,.18)--(.83,.80)--(.62,.62)--cycle;
    \fill[#4,opacity=.10] (0,0)--(.62,0)--(.62,.62)--(0,.62)--cycle;
    \draw[black!78,line width=.65pt] (0,0) rectangle ++(.62,.62);
    \draw[black!78,line width=.65pt] (.21,.18) rectangle ++(.62,.62);
    \foreach \xx/\yy in {0/0,.62/0,.62/.62,0/.62}{
      \draw[black!75,line width=.60pt] (\xx,\yy)--++(.21,.18);
    }
    \foreach \xx/\yy in {0/0,.62/0,.62/.62,0/.62}{
      \filldraw[draw=black!72,fill=#4,line width=.35pt]
        (\xx,\yy) circle[radius=.042];
      \filldraw[draw=black!72,fill=#4,line width=.35pt]
        (\xx+.21,\yy+.18) circle[radius=.042];
    }
    \node[font=\scriptsize,anchor=north] at (.43,-.07) {$C_{#3}$};
  \end{scope}%
}
\kernelcube{3.03}{1.27}{0}{oiOrange}
\kernelcube{4.11}{1.27}{1}{oiSkyBlue}
\kernelcube{5.19}{1.27}{2}{oiGreen}
\kernelcube{6.27}{1.27}{3}{oiYellow}
\kernelcube{3.03}{-.23}{4}{oiBlue}
\kernelcube{4.11}{-.23}{5}{oiVermillion}
\kernelcube{5.19}{-.23}{6}{oiPurple}
\kernelcube{6.27}{-.23}{7}{oiBlack}
\node[label] at (5.04,-1.12) {$C_0,\ldots,C_7\in\mathbb R^{2\times2\times2}$};
\end{tikzpicture}%
        }
    \end{subfigure}%
    \hfill
    \begin{subfigure}[c]{0.55\textwidth}
        \centering
        \resizebox{0.98\linewidth}{!}{%
            \begin{tikzpicture}[
  >=Latex, font=\small,
  arrow/.style={-Latex,draw=black!70,line width=.85pt},
  nodepoint/.style={
    circle,
    draw=black!65,
    fill=white,
    inner sep=0pt,
    minimum size=1.15pt
  },
  line join=round,
]

\definecolor{oiOrange}{HTML}{E69F00}
\definecolor{oiSky}{HTML}{56B4E9}
\definecolor{oiGreen}{HTML}{009E73}
\definecolor{oiYellow}{HTML}{F0E442}
\definecolor{oiBlue}{HTML}{0072B2}
\definecolor{oiVermilion}{HTML}{D55E00}
\definecolor{oiPurple}{HTML}{CC79A7}
\definecolor{oiBlack}{HTML}{444444}

\definecolor{lowerLayer}{HTML}{D4D0C9}
\definecolor{upperLayer}{HTML}{EEEAE3}
\definecolor{lowerGrid}{HTML}{736E66}
\definecolor{upperGrid}{HTML}{A49D92}

%
\newcommand{\convolutioncopy}[4]{%
  \begin{scope}[shift={(#1,#2)}]

    \begin{scope}[
      x={(.31cm,-.070cm)},
      y={(.16cm,.18cm)}
    ]
      \fill[lowerLayer,opacity=.88]
        (0,0)--(4,0)--(4,3)--(0,3)--cycle;

      \foreach \i in {0,...,4}{
        \draw[lowerGrid,line width=.4pt]
          (\i,0)--(\i,3);
      }

      \foreach \j in {0,...,3}{
        \draw[lowerGrid,line width=.4pt]
          (0,\j)--(4,\j);
      }

      \foreach \i in {0,...,4}{
        \foreach \j in {0,...,3}{
          \node[nodepoint] at (\i,\j) {};
        }
      }

      \coordinate (a) at (1,1);
      \coordinate (b) at (2,1);
      \coordinate (c) at (2,2);
      \coordinate (d) at (1,2);

    \end{scope}

    \fill[#3,opacity=.31]
      (a)--(b)--($(b)+(0,.33)$)--($(a)+(0,.33)$)--cycle;

    \fill[#3,opacity=.40]
      (b)--(c)--($(c)+(0,.33)$)--($(b)+(0,.33)$)--cycle;

    \fill[#3,opacity=.29]
      (d)--(a)--($(a)+(0,.33)$)--($(d)+(0,.33)$)--cycle;

    \foreach \p in {a,b,c,d}{
      \draw[#3!70!black,line width=.92pt]
        (\p)--($(\p)+(0,.33)$);
    }

    \draw[#3!70!black,line width=.8pt]
      (a)--(b)--(c)--(d)--cycle;

    \begin{scope}[
      shift={(0,.33)},
      x={(.31cm,-.070cm)},
      y={(.16cm,.18cm)}
    ]

      \fill[upperLayer,opacity=.50]
        (0,0)--(4,0)--(4,3)--(0,3)--cycle;

      \foreach \i in {0,...,4}{
        \draw[upperGrid,opacity=.68,line width=.4pt]
          (\i,0)--(\i,3);
      }

      \foreach \j in {0,...,3}{
        \draw[upperGrid,opacity=.68,line width=.4pt]
          (0,\j)--(4,\j);
      }

      \fill[#3,opacity=.20]
        (1,1)--(2,1)--(2,2)--(1,2)--cycle;

      \draw[#3!70!black,line width=.9pt]
        (1,1)--(2,1)--(2,2)--(1,2)--cycle;

      \foreach \i in {0,...,4}{
        \foreach \j in {0,...,3}{
          \node[nodepoint,opacity=.72] at (\i,\j) {};
        }
      }

    \end{scope}

  \end{scope}%
}

%

\convolutioncopy{0.00}{2.55}{oiOrange}{1}
\convolutioncopy{2.15}{2.55}{oiSky}{2}
\convolutioncopy{4.30}{2.55}{oiGreen}{3}
\convolutioncopy{6.45}{2.55}{oiYellow}{4}

\convolutioncopy{0.00}{0.00}{oiBlue}{5}
\convolutioncopy{2.15}{0.00}{oiVermilion}{6}
\convolutioncopy{4.30}{0.00}{oiPurple}{7}
\convolutioncopy{6.45}{0.00}{oiBlack}{8}

\draw[arrow]
  (8.55,1.52)--(9.60,1.52);

\node[font=\footnotesize,above]
  at (9.08,1.62) {$\mathcal C(U)$};

\newcommand{\outputchannel}[3]{%
  \begin{scope}[shift={(#1,#2)}]

    \fill[#3,opacity=.45]
      (0,0) rectangle (.39,.39);

    \draw[#3!70!black,line width=.9pt]
      (0,0) rectangle (.39,.39);

  \end{scope}%
}


\foreach \chan/\col [count=\k from 0] in {
  0/oiOrange,
  1/oiSky,
  2/oiGreen,
  3/oiYellow,
  4/oiBlue,
  5/oiVermilion,
  6/oiPurple,
  7/oiBlack}{%

  \pgfmathsetmacro{\cy}{1.52 + (3.5-\k)*.46}
  \pgfmathsetmacro{\by}{\cy-.195}

  \outputchannel{10.12}{\by}{\col}

  \node[
    font=\scriptsize,
    anchor=west,
    inner sep=1pt
  ] at (10.57,\cy)
    {$H_{\chan,i,j}$};

}

\end{tikzpicture}%
        }
    \end{subfigure}

    \caption{Tensor representation of the unweighted element stiffness action.
    Left: the solid-element stiffness matrix $\elementstiffness\in\mathbb{R}^{8\times8}$ reshaped into eight coefficient tensors $C_0,\ldots,C_7\in\mathbb{R}^{2\times2\times2}$.
    Right: application of these coefficients to the nodal displacement field, producing eight local force contributions per element.}

    \label{fig:matrixfree_convolution}
\end{figure}
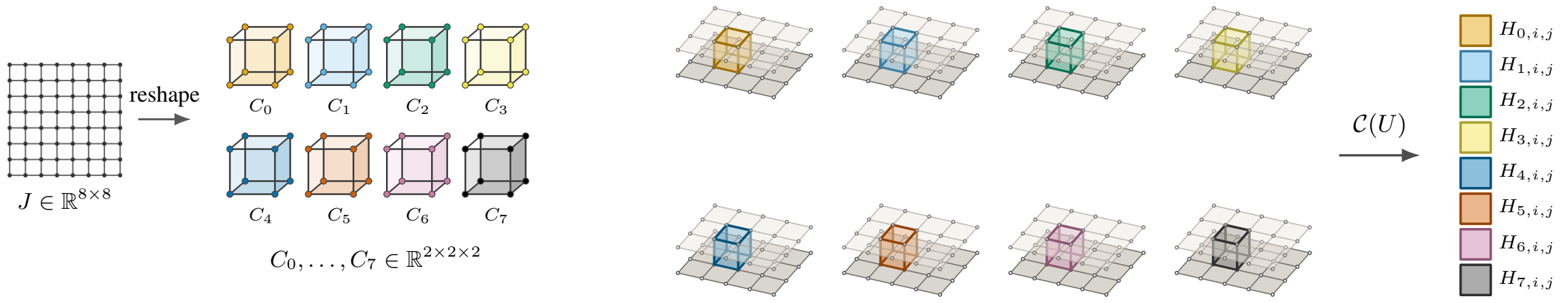

For an element $e$ indexed by $(i,j)$, the regular mesh connectivity allows each entry of its element displacement vector to be extracted directly from the nodal displacement tensor $U$ as
\begin{align}
    \left( \uvec_{e} \right)_{(4b+2\delta_x+\delta_y)}
    &=
    U_{b,i+\delta_x,j+\delta_y},
    \label{eq:matrixfree_local_displacements}
\end{align}
where $b\in\{0,1\}$ identifies the displacement component, and $\delta_x\in\{0,1\}$ and $\delta_y\in\{0,1\}$ identify one of the four nodes of element $e$.
The eight entries of $\uvec_e$ correspond to these displacements under the established local degree-of-freedom ordering.
To express the element stiffness action in tensor form, the solid-element stiffness matrix $\elementstiffness\in\mathbb{R}^{8\times8}$ is reshaped as $C\in\mathbb{R}^{8\times2\times2\times2}$.
The row index $a\in\{0,\ldots,7\}$ of $\elementstiffness$ is retained as the first index of $C$, while its column index is reshaped using the same indexing adopted for the entries of the element displacement vector: $C_{ab\delta_x\delta_y} = \elementstiffness_{a,\,4b+2\delta_x+\delta_y}$.
The unweighted element force contributions are then given by
\begin{align}
    H_{aij} &= \sum_{b=0}^{1} \sum_{\delta_x=0}^{1} \sum_{\delta_y=0}^{1} C_{ab\delta_x\delta_y} U_{b,i+\delta_x,j+\delta_y}.\label{eq:matrixfree_element_action}
\end{align}
Figure~\ref{fig:matrixfree_convolution} illustrates the tensor representation of the unweighted element stiffness action.
On the left, the solid-element stiffness matrix $\elementstiffness$ is reshaped into eight coefficient tensors $C_0,\ldots,C_7\in\mathbb{R}^{2\times2\times2}$.
On the right, these coefficients act on the nodal displacement field to produce the eight components of $H$ for each element identified via the indices $(i,j)$.

The tensor $H$ contains the unweighted element force contributions.
To account for the density-dependent material stiffness, each contribution is multiplied by the stiffness interpolation coefficient of the corresponding element,
\begin{align}
    \widehat H_{aij} &= W_{ij}H_{aij}, \label{eq:matrixfree_weighting}
\end{align}
yielding the weighted element force contributions $\widehat H\in\mathbb{R}^{8\times N_x\times N_y}$.

The final step assembles these contributions into the global nodal force field.
Since neighboring elements share nodes, each interior node receives contributions from four surrounding elements.
To express this assembly, the first dimension of $\widehat H$ is reshaped following the common local degree-of-freedom ordering of $\uvec_e$ and $\fvec_e$.
We denote the resulting tensor by
$\widehat{\mathcal H} \in \mathbb{R}^{2\times2\times2\times N_x\times N_y}$, whose entry $\widehat{\mathcal H}_{b\delta_x\delta_yij}$ contains the weighted element force contribution associated with component $b\in\{0,1\}$ and local element node $(\delta_x,\delta_y)\in\{0,1\}^2$ of the element indexed by $(i,j)$.

The local element node $(\delta_x,\delta_y)$ of the element indexed by $(i,j)$ corresponds to the global node indexed by $(m,n)$, where $m = i+\delta_x$ and $n = j+\delta_y$.
Hence, the contribution associated with local element node $(\delta_x,\delta_y)$ at the global node $(m,n)$ originates from the element indexed by $(m-\delta_x,n-\delta_y)$.
The assembled nodal force field $F\in\mathbb{R}^{2\times(N_x+1)\times(N_y+1)}$ is given by
\begin{align}
    F_{bmn} &= \sum_{\delta_x=0}^{1} \sum_{\delta_y=0}^{1} \widetilde{\mathcal H}_{ b\delta_x\delta_y,\, m-\delta_x,\, n-\delta_y }, \label{eq:matrixfree_assembly}
\end{align}
where $\widetilde{\mathcal H}$ denotes the zero extension of $\widehat{\mathcal H}$ outside the element-index domain.
This extension accounts for boundary nodes, which receive contributions from fewer than four elements.

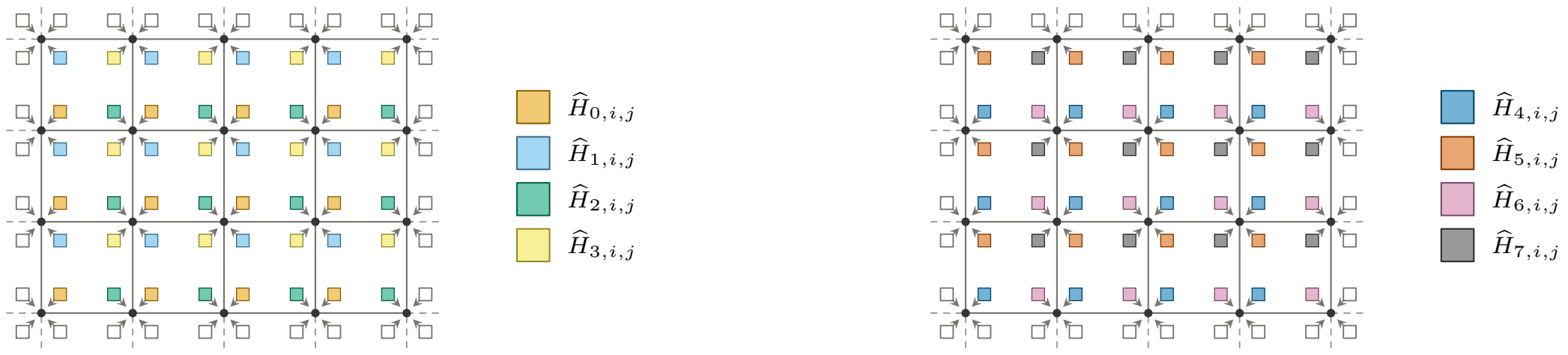
\begin{figure}[h]
    \centering

    \begin{minipage}[c]{0.49\linewidth}
        \centering

        \begin{minipage}[c]{0.5\linewidth}
            \centering
            \resizebox{\linewidth}{!}{%
\begin{tikzpicture}[line join=round]
  \definecolor{contributionInk}{HTML}{353535}
  \definecolor{oiOrange}{HTML}{E69F00}
  \definecolor{oiSky}{HTML}{56B4E9}
  \definecolor{oiGreen}{HTML}{009E73}
  \definecolor{oiYellow}{HTML}{F0E442}
  \definecolor{contributionOutline}{HTML}{77736D}
  \def\sx{1.12} 
  \def\sy{1.12} 
  \def\o{0.23}  
  \def\s{0.155} 

  \def\ext{0.43} 
  \foreach \j in {0,...,3}{
    \draw[contributionOutline,dashed,dash pattern=on 2.4pt off 2.0pt,
      line width=.48pt,opacity=.75]
      (-\ext,\j*\sy)--(0,\j*\sy);
    \draw[contributionOutline,dashed,dash pattern=on 2.4pt off 2.0pt,
      line width=.48pt,opacity=.75]
      (4*\sx,\j*\sy)--(4*\sx+\ext,\j*\sy);
  }
  \foreach \i in {0,...,4}{
    \draw[contributionOutline,dashed,dash pattern=on 2.4pt off 2.0pt,
      line width=.48pt,opacity=.75]
      (\i*\sx,-\ext)--(\i*\sx,0);
    \draw[contributionOutline,dashed,dash pattern=on 2.4pt off 2.0pt,
      line width=.48pt,opacity=.75]
      (\i*\sx,3*\sy)--(\i*\sx,3*\sy+\ext);
  }

  \foreach \j in {0,...,3}{
    \draw[contributionOutline,line width=.55pt] (0,\j*\sy) -- (4*\sx,\j*\sy);
  }
  \foreach \i in {0,...,4}{
    \draw[contributionOutline,line width=.55pt] (\i*\sx,0) -- (\i*\sx,3*\sy);
  }

  \foreach \i in {0,...,4}{
    \foreach \j in {0,...,3}{
      \foreach \dx/\dy in {-1/-1,-1/1,1/-1,1/1}{
        \pgfmathtruncatemacro{\outside}{
          (\i==0 && \dx<0) || (\i==4 && \dx>0) ||
          (\j==0 && \dy<0) || (\j==3 && \dy>0)
        }
        \pgfmathsetmacro{\cx}{\i*\sx+\dx*\o}
        \pgfmathsetmacro{\cy}{\j*\sy+\dy*\o}
        \ifnum\outside=1
          \filldraw[fill=white,draw=contributionOutline,line width=.48pt]
            (\cx-\s/2,\cy-\s/2) rectangle (\cx+\s/2,\cy+\s/2);
        \else
          \def\cornercolor{oiOrange}
          \ifnum\dx=-1
            \def\cornercolor{oiGreen}
          \fi
          \ifnum\dy=-1
            \def\cornercolor{oiSky}
            \ifnum\dx=-1
              \def\cornercolor{oiYellow}
            \fi
          \fi
          \filldraw[fill=\cornercolor!55!white,draw=\cornercolor!60!black,line width=.35pt]
            (\cx-\s/2,\cy-\s/2) rectangle (\cx+\s/2,\cy+\s/2);
        \fi
      }
      \foreach \dx/\dy in {-1/-1,-1/1,1/-1,1/1}{
        \pgfmathsetmacro{\ax}{\i*\sx+\dx*(\o-\s/2+.012)}
        \pgfmathsetmacro{\ay}{\j*\sy+\dy*(\o-\s/2+.012)}
        \pgfmathsetmacro{\bx}{\i*\sx+\dx*.077}
        \pgfmathsetmacro{\by}{\j*\sy+\dy*.077}
        \draw[->,>=stealth,draw=contributionOutline,line width=.42pt]
          (\ax,\ay)--(\bx,\by);
      }
      \fill[contributionInk] (\i*\sx,\j*\sy) circle[radius=.052cm];
    }
  }
\end{tikzpicture}%
            }
        \end{minipage}%
        \begin{minipage}[c]{0.3\linewidth}
            \centering
            \resizebox{0.5\linewidth}{!}{%
\begin{tikzpicture}[font=\small,line join=round]
  \definecolor{oiOrange}{HTML}{E69F00}
  \definecolor{oiSky}{HTML}{56B4E9}
  \definecolor{oiGreen}{HTML}{009E73}
  \definecolor{oiYellow}{HTML}{F0E442}
  \foreach \chan/\col [count=\k from 0] in {0/oiOrange,1/oiSky,2/oiGreen,3/oiYellow}{%
    \pgfmathsetmacro{\cy}{(3-\k)*.48}
    \filldraw[fill=\col!55!white,draw=\col!60!black,line width=.45pt]
      (0,\cy) rectangle (.34,\cy+.34);
    \node[font=\scriptsize,anchor=west,inner sep=1pt]
      at (.48,\cy+.17) {$\widehat{H}_{\chan,i,j}$};
  }
\end{tikzpicture}%
            }
        \end{minipage}%

    \end{minipage}%
    \hfill
    \begin{minipage}[c]{0.49\linewidth}
        \centering

        \begin{minipage}[c]{0.5\linewidth}
            \centering
            \resizebox{\linewidth}{!}{%
\begin{tikzpicture}[line join=round]
  \definecolor{contributionInk}{HTML}{353535}
  \definecolor{oiBlue}{HTML}{0072B2}
  \definecolor{oiVermilion}{HTML}{D55E00}
  \definecolor{oiPurple}{HTML}{CC79A7}
  \definecolor{oiBlack}{HTML}{444444}
  \definecolor{contributionOutline}{HTML}{77736D}
  \def\sx{1.12} 
  \def\sy{1.12} 
  \def\o{0.23}  
  \def\s{0.155} 

  \def\ext{0.43} 
  \foreach \j in {0,...,3}{
    \draw[contributionOutline,dashed,dash pattern=on 2.4pt off 2.0pt,
      line width=.48pt,opacity=.75]
      (-\ext,\j*\sy)--(0,\j*\sy);
    \draw[contributionOutline,dashed,dash pattern=on 2.4pt off 2.0pt,
      line width=.48pt,opacity=.75]
      (4*\sx,\j*\sy)--(4*\sx+\ext,\j*\sy);
  }
  \foreach \i in {0,...,4}{
    \draw[contributionOutline,dashed,dash pattern=on 2.4pt off 2.0pt,
      line width=.48pt,opacity=.75]
      (\i*\sx,-\ext)--(\i*\sx,0);
    \draw[contributionOutline,dashed,dash pattern=on 2.4pt off 2.0pt,
      line width=.48pt,opacity=.75]
      (\i*\sx,3*\sy)--(\i*\sx,3*\sy+\ext);
  }

  \foreach \j in {0,...,3}{
    \draw[contributionOutline,line width=.55pt] (0,\j*\sy) -- (4*\sx,\j*\sy);
  }
  \foreach \i in {0,...,4}{
    \draw[contributionOutline,line width=.55pt] (\i*\sx,0) -- (\i*\sx,3*\sy);
  }

  \foreach \i in {0,...,4}{
    \foreach \j in {0,...,3}{
      \foreach \dx/\dy in {-1/-1,-1/1,1/-1,1/1}{
        \pgfmathtruncatemacro{\outside}{
          (\i==0 && \dx<0) || (\i==4 && \dx>0) ||
          (\j==0 && \dy<0) || (\j==3 && \dy>0)
        }
        \pgfmathsetmacro{\cx}{\i*\sx+\dx*\o}
        \pgfmathsetmacro{\cy}{\j*\sy+\dy*\o}
        \ifnum\outside=1
          \filldraw[fill=white,draw=contributionOutline,line width=.48pt]
            (\cx-\s/2,\cy-\s/2) rectangle (\cx+\s/2,\cy+\s/2);
        \else
          \def\cornercolor{oiBlue}
          \ifnum\dx=-1
            \def\cornercolor{oiPurple}
          \fi
          \ifnum\dy=-1
            \def\cornercolor{oiVermilion}
            \ifnum\dx=-1
              \def\cornercolor{oiBlack}
            \fi
          \fi
          \filldraw[fill=\cornercolor!55!white,draw=\cornercolor!60!black,line width=.35pt]
            (\cx-\s/2,\cy-\s/2) rectangle (\cx+\s/2,\cy+\s/2);
        \fi
      }
      \foreach \dx/\dy in {-1/-1,-1/1,1/-1,1/1}{
        \pgfmathsetmacro{\ax}{\i*\sx+\dx*(\o-\s/2+.012)}
        \pgfmathsetmacro{\ay}{\j*\sy+\dy*(\o-\s/2+.012)}
        \pgfmathsetmacro{\bx}{\i*\sx+\dx*.077}
        \pgfmathsetmacro{\by}{\j*\sy+\dy*.077}
        \draw[->,>=stealth,draw=contributionOutline,line width=.42pt]
          (\ax,\ay)--(\bx,\by);
      }
      \fill[contributionInk] (\i*\sx,\j*\sy) circle[radius=.052cm];
    }
  }
\end{tikzpicture}%
            }
        \end{minipage}%
        \begin{minipage}[c]{0.3\linewidth}
            \centering
            \resizebox{0.5\linewidth}{!}{%
\begin{tikzpicture}[font=\small,line join=round]
  \definecolor{oiBlue}{HTML}{0072B2}
  \definecolor{oiVermilion}{HTML}{D55E00}
  \definecolor{oiPurple}{HTML}{CC79A7}
  \definecolor{oiBlack}{HTML}{444444}
  \foreach \chan/\col [count=\k from 0] in {4/oiBlue,5/oiVermilion,6/oiPurple,7/oiBlack}{%
    \pgfmathsetmacro{\cy}{(3-\k)*.48}
    \filldraw[fill=\col!55!white,draw=\col!60!black,line width=.45pt]
      (0,\cy) rectangle (.34,\cy+.34);
    \node[font=\scriptsize,anchor=west,inner sep=1pt]
      at (.48,\cy+.17) {$\widehat{H}_{\chan,i,j}$};
  }
\end{tikzpicture}%
            }
        \end{minipage}

    \end{minipage}

    \caption{Assembly of weighted element force contributions into the nodal force field.
    Left: assembly of the $x$-component and its corresponding color key.
    Right: assembly of the $y$-component and its corresponding color key.
    Colored squares represent contributions from elements within the mesh, while white squares represent zero-valued contributions introduced by padding.
    Arrows indicate the accumulation of contributions at the corresponding nodes.}

    \label{fig:matrixfree_assembly}
\end{figure}

Figure~\ref{fig:matrixfree_assembly} illustrates this assembly for both displacement components.
Each node receives four local force contributions, represented by the squares surrounding it.
At interior nodes, all four contributions originate from elements within the mesh.
At boundary nodes, the missing contributions are replaced by zeros, represented by the white squares.
The arrows indicate how these contributions accumulate at their corresponding nodal positions.

We denote the tensor operation defined in Equation~\ref{eq:matrixfree_element_action} by $\mathcal T$ and the assembly operation defined in Equation~\ref{eq:matrixfree_assembly} by $\mathcal A$.
Combining these operations with the element-wise stiffness interpolation gives the complete matrix-free stiffness action,
\begin{align}
    \stiffness\uvec
    &=
    \operatorname{vec}\left(
        \mathcal A\left(
            W\odot\mathcal T(U)
        \right)
    \right),
    \label{eq:matrixfree_operator}
\end{align}
where $\odot$ denotes element-wise multiplication with broadcasting over the eight local force contributions, and $\operatorname{vec}$ maps the assembled nodal force tensor to the global force vector.

\subsection{Convolution-based matrix-free operator}

The tensor operation $\mathcal T$ derived in the preceding subsection admits a particularly efficient computational realization.
Each component $H_{aij}$ depends on the displacements at the four nodes of the element identified by $(i,j)$, with the same coefficients $C_{ab\delta_x\delta_y}$ applied throughout the mesh.
This operation has the structure of a two-dimensional convolution with two input channels, eight output channels, and a spatial kernel size of $2\times2$.

The two input channels correspond to the nodal displacement fields $U_x$ and $U_y$, which together form the displacement tensor $U$.
The eight output channels correspond to the eight components of the unweighted element force vector, with each channel containing one component for every element in the mesh.
The coefficient tensor $C\in\mathbb{R}^{8\times2\times2\times2}$ provides the convolution kernel weights.
Applying the convolution to $U\in\mathbb{R}^{2\times(N_x+1)\times(N_y+1)}$ with unit stride and no padding produces $H\in\mathbb{R}^{8\times N_x\times N_y}$.

This interpretation allows us to replace the repeated element-wise matrix--vector products with a single convolution operation.
Combined with element-wise weighting and structured assembly, the resulting formulation reproduces the action of the assembled finite-element stiffness matrix without constructing the global sparse matrix or storing element connectivity arrays.
By expressing the stiffness action through standard computational primitives provided by high-performance numerical libraries, such as JAX, we can leverage hardware-optimized implementations across different architectures, including SIMD instructions on CPUs and specialized GPU kernels, without developing custom low-level code.

Listing~\ref{lst:matrixfree2d} presents a compact implementation using JAX~\citep{jax2018github} and Einops~\citep{rogozhnikov2022einops}.
The nodal displacement tensor $U$ and the element-wise stiffness coefficients $W$ correspond to the arrays \mintinline{python}{u} and \mintinline{python}{w}, respectively.
The solid-element stiffness matrix $\elementstiffness$ is supplied as \mintinline{python}{J} and reshaped into the convolution kernel $C\in\mathbb{R}^{8\times2\times2\times2}$ in line~\ref{algjax_matrixfree:line:kernel}.
The JAX primitive \mintinline{python}{lax.conv_general_dilated} evaluates Equation~\ref{eq:matrixfree_element_action}, producing the array \mintinline{python}{h} of unweighted element force contributions.
The subsequent multiplication by \mintinline{python}{w} introduces the density-dependent stiffness interpolation coefficients, as defined by Equation~\ref{eq:matrixfree_weighting}.
Finally, the combination of \mintinline{python}{rearrange}, \mintinline{python}{jnp.pad}, and \mintinline{python}{reduce} implements the assembly operation in Equation~\ref{eq:matrixfree_assembly}.
The \mintinline{python}{rearrange} operation places the local force contributions at the corresponding element nodes.
Padding extends the resulting tensor to account for contributions outside the computational domain, while the reduction sums the contributions associated with each global node.
The resulting array \mintinline{python}{f} represents the assembled nodal force field $F\in\mathbb{R}^{2\times(N_x+1)\times(N_y+1)}$.

\begin{listing}[t]
\begin{minted}[
    fontsize=\footnotesize,
    linenos,
    breaklines,
    escapeinside=||
]{python}
import jax.numpy as jnp
from jax import lax
from einops import rearrange, reduce

def matrix_free_2d(u, J, w):
    # Construct kernel from the solid-element stiffness matrix
    kernel = J.reshape(8, 2, 2, 2) |\label{algjax_matrixfree:line:kernel}|

    # Evaluate the unweighted element stiffness actions
    h = lax.conv_general_dilated(
        u[None, ...],
        kernel,
        window_strides=(1, 1),
        padding="VALID",
        dimension_numbers=("NCHW", "OIHW", "NCHW"),
    )[0] |\label{algjax_matrixfree:line:conv}|

    # Apply density-dependent stiffness coefficients
    h_weighted = w * h |\label{algjax_matrixfree:line:weighting}|

    # Distribute contributions to the element nodes
    h_weighted = rearrange(h_weighted, "(c i j) x y -> c (x i) (y j)", c=2, i=2, j=2) |\label{algjax_matrixfree:line:rearrange}|

    # Assemble contributions at shared nodes
    h_weighted = jnp.pad(h_weighted, ((0, 0), (1, 1), (1, 1))) |\label{algjax_matrixfree:line:pad}|
    f = reduce(h_weighted, "c (x i) (y j) -> c x y", "sum", i=2, j=2) |\label{algjax_matrixfree:line:reduce}|

    return f
\end{minted}
\caption{Convolution-based matrix-free finite-element operator in two dimensions using JAX primitives.}
\label{lst:matrixfree2d}
\end{listing}

\subsection{Handling boundary conditions}

In finite-element analysis, Dirichlet boundary conditions prescribe the displacement at selected degrees of freedom.
In the context of topology optimization, these conditions typically represent supports that prevent displacement in one or more directions.
The remaining, unconstrained degrees of freedom are determined by solving the equilibrium equations.

For homogeneous Dirichlet boundary conditions, the prescribed displacements vanish.
We can therefore restrict the equilibrium equations to the unconstrained degrees of freedom.
Let $N_{\mathrm{dof}}$ denote the total number of degrees of freedom and $N_{\mathrm{free}}$ the number of unconstrained degrees of freedom.
Introducing the restriction matrix $R\in\mathbb{R}^{N_{\mathrm{free}}\times N_{\mathrm{dof}}}$, which extracts the unconstrained entries of a vector, the reduced stiffness operator takes the form
\begin{align}
    \stiffness_f
    &=
    R\stiffness R^{\mathsf T}.
    \label{eq:matrixfree_restricted_operator}
\end{align}

The restriction matrix $R$ provides a reduced-space representation in which the constrained degrees of freedom are removed and only the $N_{\mathrm{free}}$ unconstrained degrees of freedom are retained.
Although the reduced-space formulation is standard, explicitly removing the constrained degrees of freedom would disrupt the regular full-grid tensor layout required by the convolution-based matrix-free operation.
We therefore retain all degrees of freedom in the tensor representation and enforce the homogeneous constraints using a binary mask $M\in\{0,1\}^{2\times(N_x+1)\times(N_y+1)}$.
The entries of $M$ are equal to one at unconstrained degrees of freedom and zero at constrained degrees of freedom.
Applying the mask before and after the stiffness operation gives
\begin{align}
    F &= M\odot \mathcal A\left( W\odot\mathcal T(M\odot U) \right), \label{eq:matrixfree_masked_operator} 
\end{align}
where the inner mask enforces zero displacement at the constrained degrees of freedom, while the outer mask removes the corresponding force entries after assembly.

Unlike the reduced-space formulation, the masked formulation retains all $N_{\mathrm{dof}}$ degrees of freedom in the displacement and force representations.
The constrained entries, however, remain equal to zero and do not participate in the iterative solution.
When restricted to the unconstrained subspace, the masked stiffness action is equivalent to the reduced operator $R\stiffness R^{\transpose}$.

In the full space, the masked stiffness operator has zero rows and columns at the constrained degrees of freedom and is therefore singular.
The conjugate gradient method is consequently initialized and applied within the unconstrained subspace by keeping the constrained entries of the solution, right-hand side, residuals, and search directions equal to zero.
Listing~\ref{lst:matrixfree2d_dirichlet} shows the corresponding JAX implementation.
The binary mask is represented by the array \mintinline{python}{free}, and the homogeneous constraints are enforced using only two element-wise multiplications.

\begin{listing}[t]
\begin{minted}[
    fontsize=\footnotesize,
    linenos,
    breaklines
]{python}
def matrix_free_2d_dirichlet(u, J, w, free):
    u = free * u
    f = matrix_free_2d(u, J, w)
    return free * f
\end{minted}
\caption{Application of the matrix-free stiffness operator with homogeneous Dirichlet boundary conditions using a binary mask of unconstrained degrees of freedom.}
\label{lst:matrixfree2d_dirichlet}
\end{listing}

\subsection{Slicing the matrix-free operator}
\label{sec:slicing_matrix_free}

Although the convolution-based formulation avoids assembling and storing the global stiffness matrix, evaluating the operator over the entire domain produces intermediate tensors whose sizes scale with the number of elements.
In particular, the tensor of element force contributions contains eight entries per element in two dimensions and twenty-four entries per element in three dimensions.
To limit the memory required by these intermediate quantities, we partition the nodal force field into slices and evaluate them separately.

Consider a two-dimensional mesh with $N_x\times N_y$ elements.
We divide the nodal positions along the $x$-direction into slices of $B_\mathrm{slice}$ consecutive planes, each spanning the full extent of the domain along the $y$-direction.
Assuming that $N_x+1$ is divisible by $B_\mathrm{slice}$, we index the slices by $\sigma \in \{0,\ldots,(N_x+1)/B_\mathrm{slice}-1\}$.
Slice $\sigma$ contains the nodal planes indexed by
\begin{align}
    \nu \in \{\sigma B_\mathrm{slice},\ldots,(\sigma+1)B_\mathrm{slice}-1\}.
\end{align}

The locality of the finite-element stiffness action allows each slice to be evaluated using only a small overlap with its neighboring slices.
The nodal forces at a given position depend only on contributions from the immediately adjacent elements.
Consequently, computing the nodal forces within slice $\sigma$ requires element contributions from planes indexed by
\begin{align}
    \epsilon \in \{\sigma B_\mathrm{slice}-1,\ldots,(\sigma+1)B_\mathrm{slice}-1\}.
\end{align}
These elements, in turn, depend on nodal displacements at planes indexed by
\begin{align}
    \nu \in \{\sigma B_\mathrm{slice}-1,\ldots,(\sigma+1)B_\mathrm{slice}\}.
\end{align}

Thus, evaluating $B_\mathrm{slice}$ output planes requires $B_\mathrm{slice}+2$ planes of nodal displacements and $B_\mathrm{slice}+1$ planes of element stiffness coefficients.
The additional planes provide the overlap needed to evaluate the stiffness action at the slice boundaries.
At the domain boundaries, the required overlap is supplied by zero padding, consistent with the assembly operation introduced previously.
Only the $B_\mathrm{slice}$ assigned output planes are retained, so the resulting slices do not overlap and can be concatenated to recover the complete nodal force field.

Listing~\ref{lst:matrixfree2d_sliced} illustrates the slicing procedure using the matrix-free operator introduced in Listing~\ref{lst:matrixfree2d}.
The nodal displacement and element-wise stiffness tensors are first padded along the slicing direction.
For each slice, we extract the overlapping input regions, evaluate the matrix-free operator, and retain only the assigned nodal force planes.
Concatenating these outputs recovers the complete nodal force field.

\begin{listing}[t]
\begin{minted}[
    fontsize=\footnotesize,
    linenos,
    breaklines
]{python}
import jax.numpy as jnp

def matrix_free_2d_sliced(u, J, w, B_slice):
    # u: (2, Nx+1, Ny+1)
    # w: (Nx, Ny)

    # Pad the input tensors along x
    u_pad = jnp.pad(u, ((0, 0), (1, 1), (0, 0)))
    w_pad = jnp.pad(w, ((1, 1), (0, 0)))

    slices = []

    for sigma in range(u.shape[1] // B_slice):
        i = sigma * B_slice

        # Extract overlapping input regions
        u_local = u_pad[:, i:i+B_slice+2, :]                 # (2, B_slice+2, Ny+1)
        w_local = w_pad[i:i+B_slice+1, :]                    # (B_slice+1, Ny)

        # Evaluate the matrix-free operator
        f_local = matrix_free_2d(u_local, J, w_local)  # (2, B_slice+2, Ny+1)

        # Retain the assigned output planes
        slices.append(f_local[:, 1:-1, :])             # (2, B_slice, Ny+1)

    # Recover the complete nodal force field
    return jnp.concatenate(slices, axis=1)             # (2, Nx+1, Ny+1)
\end{minted}
\caption{Simplified implementation of the sliced matrix-free operator in two dimensions.
The domain is divided into slices of $B_{\mathrm{slice}}$ nodal planes along the $x$-direction.
The tensor shapes are indicated in the comments.}
\label{lst:matrixfree2d_sliced}
\end{listing}

While Listing~\ref{lst:matrixfree2d_sliced} illustrates the slicing procedure, its implementation is not yet optimized for GPU execution.
In particular, collecting the output slices in a Python list requires the intermediate results to be retained until concatenation.
A more efficient implementation replaces the Python loop with JAX control-flow primitives, such as \mintinline{python}{lax.scan} or \mintinline{python}{lax.map}.
These primitives express the iteration within the compiled computation, allowing the slices to be processed without unrolling the loop into a large sequence of operations.
Together with dynamic slicing, they enable the matrix-free operator to process large domains while limiting the size of the intermediate tensors.

To estimate the memory savings, consider a mesh with $N_x\times N_y$ elements and a slice size $B_{\mathrm{slice}}$.
Evaluating the matrix-free operator over the entire domain produces intermediate tensors with $\mathcal O(N_xN_y)$ entries.
In contrast, evaluating one slice requires intermediate tensors with only $\mathcal O(B_{\mathrm{slice}}N_y)$ entries.
The peak memory required for these intermediate quantities is therefore reduced by approximately a factor of $N_x/B_{\mathrm{slice}}$, assuming that the slices are processed sequentially.
This reduction becomes particularly significant for large meshes when $B_{\mathrm{slice}}\ll N_x$.

Although GPU programming generally favors parallel execution, processing the slices sequentially does not necessarily introduce a significant runtime penalty.
For sufficiently large problems, each slice contains enough elements to expose substantial parallelism within the convolution and tensor operations.
Moreover, the matrix-free stiffness action involves relatively little computation compared with the amount of data accessed, making its execution largely limited by memory bandwidth rather than arithmetic throughput.
Consequently, processing the slices sequentially can maintain high GPU utilization, provided that the slices are sufficiently large.
The reduction in intermediate memory requirements can therefore be achieved with little impact on the overall operator evaluation time.

\subsection{Extension to three dimensions}

The matrix-free formulation extends naturally to structured meshes of trilinear hexahedral elements.
Each element has eight nodes with three displacement degrees of freedom per node, giving a total of twenty-four local degrees of freedom.
Accordingly, the solid-element stiffness matrix satisfies $\elementstiffness\in\mathbb{R}^{24\times24}$.
For a mesh comprising $N_x\times N_y\times N_z$ elements, we represent the nodal displacement field by the tensor $U\in\mathbb{R}^{3\times(N_x+1)\times(N_y+1)\times(N_z+1)}$, while the element-wise stiffness coefficients form $W\in\mathbb{R}^{N_x\times N_y\times N_z}$.

Following the component-first ordering introduced for the two-dimensional case, the three displacement components are ordered first by component and then by the local element nodes.
Within each component, the eight nodes are ordered first along the $x$-direction, then along the $y$-direction, and finally along the $z$-direction.
The solid-element stiffness matrix is reshaped as $C\in\mathbb{R}^{24\times3\times2\times2\times2}$.
The unweighted element force contributions then satisfy
\begin{align}
    H_{aijk}
    &=
    \sum_{b=0}^{2}
    \sum_{\delta_x=0}^{1}
    \sum_{\delta_y=0}^{1}
    \sum_{\delta_z=0}^{1}
    C_{ab\delta_x\delta_y\delta_z}
    U_{b,i+\delta_x,j+\delta_y,k+\delta_z},
    \label{eq:matrixfree_element_action_3d}
\end{align}
where $b\in\{0,1,2\}$ identifies the displacement component and $a\in\{0,\ldots,23\}$ follows the common local degree-of-freedom ordering adopted for $\uvec_e$ and $\fvec_e$.

As in two dimensions, this operation corresponds to a convolution, now with three input channels, twenty-four output channels, and a spatial kernel size of $2\times2\times2$.
The extension requires only minor modifications to the JAX implementation, as illustrated in Listing~\ref{lst:matrixfree3d}.
With unit strides and no padding, the convolution produces one set of local force contributions per element.

\begin{listing}[h]
\begin{minted}[
    fontsize=\footnotesize,
    linenos,
    breaklines
]{python}
# J: (24, 24)
kernel = J.reshape(24, 3, 2, 2, 2)

# u: (3, Nx+1, Ny+1, Nz+1)
h = lax.conv_general_dilated(
    u[None, ...],
    kernel,
    window_strides=(1, 1, 1),
    padding="VALID",
    dimension_numbers=("NCDHW", "OIDHW", "NCDHW"),
)[0]
# h: (24, Nx, Ny, Nz)
\end{minted}
\caption{Three-dimensional convolution-based element stiffness action using JAX primitives.}
\label{lst:matrixfree3d}
\end{listing}

The resulting tensor $H\in\mathbb{R}^{24\times N_x\times N_y\times N_z}$ contains the unweighted element force contributions.
As in two dimensions, these contributions are multiplied by the element-wise stiffness coefficients before being assembled into the global nodal force field.
The assembly follows the same indexing principle, with each interior node receiving contributions from eight surrounding elements.

The slicing strategy extends directly to three dimensions.
We divide the domain into slabs along one spatial direction, while retaining its full extent in the other two directions.
For example, when slicing along the $x$-direction, a slab responsible for $B_{\mathrm{slice}}$ nodal planes requires $B_{\mathrm{slice}}+2$ displacement planes and $B_{\mathrm{slice}}+1$ element-weight planes.
The local operator is evaluated on these overlapping inputs, and only the $B_{\mathrm{slice}}$ assigned output planes are retained.
Concatenating the resulting slabs recovers the complete nodal force field.

Slicing becomes particularly beneficial in three dimensions, where the intermediate element force tensor contains twenty-four entries per element.
For a mesh with $N_x\times N_y\times N_z$ elements, the unsliced evaluation requires intermediate tensors with $\mathcal O(N_xN_yN_z)$ entries.
Slicing along the $x$-direction reduces the corresponding per-slab requirement to $\mathcal O(B_{\mathrm{slice}}N_yN_z)$, while preserving the same matrix-free stiffness action.


\section{Slicing Galerkin operators}
\label{sec:galerkin_slicing}

\subsection{Element-wise Galerkin decomposition}

Consider an element $e$ at level $\ell$ of the multigrid hierarchy and let $\uvec_e$ denote the displacement vector associated with the four nodes of element $e$.
Under dyadic coarsening, each element at level $\ell$ contains $2^\ell$ fine-grid elements along each spatial direction, giving a total of $4^\ell$ fine-grid elements in two dimensions.
We index these fine elements by $s\in \{0,\ldots,4^\ell-1 \}$.
For each fine element $s$, let $Q_s^{(\ell)}$ denote the interpolation matrix that evaluates the coarse-element displacement field at the nodes of the fine element $s$.
The resulting fine-element displacement vector is
\begin{align}
    \uvec_s
    &=
    Q_s^{(\ell)}\uvec_e.\label{eq:u_s}
\end{align}
The elastic energy associated with the fine elements contained within $e$ is given by~\citep{bathe2014finite}
\begin{align}
    \mathcal E_e
    &=
    \frac{1}{2}
    \sum_{s=0}^{4^\ell-1}
    \uvec_s^{\mathsf T}
    \stiffness_s
    \uvec_s.\label{eq:galerkin_fine_energy}
\end{align}
Substituting Equation~\ref{eq:u_s} into Equation~\ref{eq:galerkin_fine_energy} yields
\begin{align}
    \mathcal E_e
    =
    \frac{1}{2}
    \sum_{s=0}^{4^\ell-1}
    (Q_s^{(\ell)}\uvec_e)^{\mathsf T}
    \stiffness_s
    (Q_s^{(\ell)}\uvec_e)
    =
    \frac{1}{2}
    \uvec_e^{\mathsf T}
    \left(
        \sum_{s=0}^{4^\ell-1}
        (Q_s^{(\ell)})^{\mathsf T}
        \stiffness_s
        Q_s^{(\ell)}
    \right)
    \uvec_e.
    \label{eq:galerkin_interpolated_energy}
\end{align}

Consistent with the element-agglomeration Galerkin construction of \citet{jones2001amge}, the $\ell$-level stiffness matrix $\stiffness_e^{(\ell)}$ is defined such that its quadratic form reproduces the elastic energy of the interpolated fine-element displacement field: $\mathcal E_e=\frac{1}{2}\uvec_e^{\mathsf T} \stiffness_e^{(\ell)}\uvec_e$.
Equating this expression with Equation~\ref{eq:galerkin_interpolated_energy} for every $\uvec_e$ gives
\begin{align}
    \stiffness_e^{(\ell)}
    =
    \sum_{s=0}^{4^\ell-1}
    (Q_s^{(\ell)})^{\mathsf T}
    \stiffness_s
    Q_s^{(\ell)}
    =
    \sum_{s=0}^{4^\ell-1}
    w_{e,s}
    (Q_s^{(\ell)})^{\mathsf T}
    \elementstiffness
    Q_s^{(\ell)},
    \label{eq:galerkin_coarse_element}
\end{align}
where the second equality follows from $\stiffness_s=w_{e,s}\elementstiffness$, and $w_{e,s}$ is the stiffness interpolation coefficient of the fine element $s$ contained within element $e$.

The geometric and interpolation-dependent part of each fine-element contribution is independent of the material distribution.
We therefore introduce the fixed level-dependent matrices
$
    G_s^{(\ell)} \coloneq \left(Q_s^{(\ell)}\right)^{\mathsf T} \elementstiffness Q_s^{(\ell)}.
$
The $\ell$-level element stiffness matrix, \cf Equation~\ref{eq:galerkin_coarse_element}, can then be written as
\begin{align}
    \stiffness_e^{(\ell)}
    =
    \sum_{s=0}^{4^\ell-1} w_{e,s}G_s^{(\ell)}.
    \label{eq:galerkin_weighted_operators}
\end{align}
At the finest level, $\ell=0$, element $e$ contains only itself.
In this case, $Q_0^{(0)}$ is the identity matrix, $G_0^{(0)}=\elementstiffness$, and we identify $w_{e,0}$ with $w_e$.
Equation~\ref{eq:galerkin_weighted_operators} therefore reduces to $\stiffness_e^{(0)} = w_e\elementstiffness = \stiffness_e$, which recovers the fine-grid SIMP interpolation in Equation~\ref{eq:homogeneous_Ke}.
The fine-grid formulation is thus the level-$0$ instance of the Galerkin representation.
At coarser levels, each element stiffness matrix combines $4^\ell$ fixed local operators with the corresponding fine-grid stiffness interpolation coefficients.

The element-wise decomposition of Galerkin operators is well established in the multigrid literature, particularly in approaches based on element agglomeration~\citep{jones2001amge}.
Previous work has also demonstrated matrix-free multigrid methods on GPUs, including applications to finite-element analysis and high-resolution topology optimization~\citep{dick2011real,wu2015system}.
Here, we exploit this decomposition to evaluate the Galerkin stiffness action directly from the original fine-grid stiffness interpolation coefficients.
Instead of explicitly evaluating and storing $\stiffness_e^{(\ell)}$ for every coarse element, we retain its factorized representation and apply the precomputed matrices $G_s^{(\ell)}$ using the corresponding weights $w_{e,s}$.
This avoids materializing the coarse-element stiffness matrices while retaining the Galerkin relationship between grid levels.
In the following subsection, we show how this operation can be implemented through tensor transformations and contractions.

\subsection{Matrix-free evaluation of Galerkin operators}

The decomposition derived in the preceding subsection allows us to evaluate the Galerkin stiffness action without explicitly constructing individual coarse-element stiffness matrices per element.
Instead, we retain the precomputed operators $G_s^{(\ell)}$ and combine their actions using the corresponding fine-grid stiffness interpolation coefficients.

Consider a Cartesian mesh at level $\ell$, containing $N_x^{(\ell)}\times N_y^{(\ell)}$ elements.
We collect the fine-grid stiffness interpolation coefficients associated with each coarse element into the tensor $W^{(\ell)} \in \mathbb{R}^{N_x^{(\ell)}\times N_y^{(\ell)}\times4^\ell}$, where $W^{(\ell)}_{ijs}$ denotes the coefficient of fine subelement $s$ within coarse element $(i,j)$.
This tensor is obtained by rearranging the original fine-grid stiffness interpolation coefficients, without averaging or otherwise modifying their values.
The nodal displacement field at level $\ell$ is represented by $U^{(\ell)}\in \mathbb{R}^{2\times(N_x^{(\ell)}+1)\times(N_y^{(\ell)}+1)}$.
As in the fine-grid formulation, the regular mesh connectivity allows the element displacement vectors to be extracted directly from the nodal tensor.
Collecting these vectors gives $ E^{(\ell)} \in\mathbb{R}^{8\times N_x^{(\ell)}\times N_y^{(\ell)}}$, where $E^{(\ell)}_{qij}$ is the $q$-th entry of the displacement vector of the element indexed by $(i,j)$.
The local Galerkin stiffness action is then given by
\begin{align}
    H^{(\ell)}_{aij}
    &=
    \sum_{s=0}^{4^\ell-1} W^{(\ell)}_{ijs} \sum_{q=0}^{7} G^{(\ell)}_{saq} E^{(\ell)}_{qij},
    \label{eq:galerkin_matrixfree_action}
\end{align}
where $a \in \{0,\ldots,7\}$ and $q \in \{0,\ldots,7\}$ index the output and input entries, respectively, following the common local degree-of-freedom ordering, and $G^{(\ell)}_{saq}=\left(G_s^{(\ell)}\right)_{aq}$.
The tensor $H^{(\ell)}\in \mathbb{R}^{8\times N_x^{(\ell)}\times N_y^{(\ell)}}$ contains the resulting element force contributions.

Unlike the fine-grid formulation, which applies a single reference-element matrix followed by scalar weighting, Equation~\ref{eq:galerkin_matrixfree_action} combines several fixed operators within each element.
Nevertheless, the calculation consists entirely of regular tensor operations.
In particular, the weighted matrix--vector products and their summation can be evaluated through a single tensor contraction.

Listing~\ref{lst:galerkin_matrixfree2d} illustrates the procedure using again JAX and Einops primitives.
The precomputed Galerkin matrices are supplied as \mintinline{python}{G}, while \mintinline{python}{w} contains the original fine-grid stiffness interpolation coefficients.
The \mintinline{python}{rearrange} operation groups these coefficients by coarse element and subelement.
In Line~\ref{lst:galerkin_matrixfree2d:line:conv} we use a JAX primitive to compute a convolution extracting the local displacement vectors and Line~\ref{algjax_galerkin:line:contraction} evaluates Equation~\ref{eq:galerkin_matrixfree_action}.
The resulting local force contributions are assembled using the same tensor transformations introduced for the fine-grid operator.
Consequently, the Galerkin stiffness action requires neither the construction of a global sparse matrix nor the explicit materialization of individual coarse-element stiffness matrices.

\begin{listing}[t]
\begin{minted}[
    fontsize=\footnotesize,
    linenos,
    breaklines,
    escapeinside=||
]{python}
import jax.numpy as jnp
from jax import lax
from einops import rearrange, reduce, einsum

def galerkin_matrix_free_2d(u, G, w, l):
    # u: (2, Mx+1, My+1)
    # G: (4**l, 8, 8)
    # w: (Nx, Ny)
    # l: level number (integer)

    r = 2**l
    Mx, My = u.shape[1] - 1, u.shape[2] - 1

    # Group fine-grid coefficients by coarse element
    weights = rearrange(w, "(x i) (y j) -> (x y) (i j)", i=r, j=r)      # (Mx*My, 4**l)

    # Extract local coarse-element displacements
    patches = lax.conv_general_dilated_patches(|\label{lst:galerkin_matrixfree2d:line:conv}|
        u[None, ...],
        filter_shape=(2, 2),
        window_strides=(1, 1),
        padding="VALID",
        dimension_numbers=("NCHW", "OIHW", "NCHW"),
    )[0]
    local_u = rearrange(patches, "a x y -> (x y) a")                    # (Mx*My, 8)

    # Apply weighted Galerkin operators
    local_f = einsum(weights, G, local_u, "e s, s a q, e q -> e a")     # (Mx*My, 8) |\label{algjax_galerkin:line:contraction}|

    # Assemble local contributions at shared nodes
    local_f = rearrange( |\label{algjax_galerkin:line:assembly}|
        local_f,
        "(x y) (c i j) -> c (x i) (y j)",
        x=Mx, y=My, c=2, i=2, j=2
    )
    local_f = jnp.pad(local_f, ((0, 0), (1, 1), (1, 1)))
    f = reduce(local_f, "c (x i) (y j) -> c x y", "sum", i=2, j=2)       # (2, Mx+1, My+1)

    return f  
\end{minted}
\caption{Simplified matrix-free application of the two-dimensional Galerkin stiffness operator using precomputed local matrices and fine-grid stiffness interpolation coefficients.}
\label{lst:galerkin_matrixfree2d}
\end{listing}

\subsection{Batching and Slicing the Galerkin operator}
\label{sec:sliced_galerkin}

Listing~\ref{lst:galerkin_matrixfree2d} evaluates the tensor contraction over all coarse elements simultaneously.
For large problems, however, the intermediate arrays can impose substantial memory requirements.
We therefore divide the coarse elements into smaller batches and evaluate their local stiffness actions sequentially.
Unlike the spatial slicing introduced for the fine-grid operator, this decomposition requires no overlapping input regions, since the local force contributions of different coarse elements can be computed independently.

For a two-dimensional mesh containing $N_x^{(\ell)}\times N_y^{(\ell)}$ coarse elements, the complete contraction involves $4^\ell \times N_x^{(\ell)}\times N_y^{(\ell)}$ subelement contributions.
Processing $B_{\mathrm{batch}}$ coarse elements at a time reduces the number of contributions evaluated in a single batch to $B_\mathrm{batch}\times 4^\ell$.
The total computational work remains unchanged, but the number of subelement contributions processed simultaneously is reduced by approximately a factor of $(N_x^{(\ell)} \times N_y^{(\ell)})/B_\mathrm{batch}$.
This limits the temporary storage associated with the contraction, although the input coefficients and resulting force contributions must still be stored.

Listing~\ref{lst:galerkin_sliced} illustrates the batched evaluation using \mintinline{python}{lax.map}.
The code replaces the tensor contraction at line~\ref{algjax_galerkin:line:contraction} of Listing~\ref{lst:galerkin_matrixfree2d}, while leaving the subsequent assembly unchanged.
The function \mintinline{python}{apply_element} computes the weighted Galerkin stiffness action for a single coarse element, while \mintinline{python}{lax.map} evaluates this function over batches of at most $B_\mathrm{batch}$ elements.
This allows the contraction to be executed without unrolling the complete computation.

\begin{listing}[t]
\begin{minted}[
    fontsize=\footnotesize,
    linenos,
    breaklines,
    escapeinside=||
]{python}
def apply_element(inputs):
    w_e, u_e = inputs
    return einsum(w_e, G, u_e, "s, s a q, q -> a")

# Evaluate local stiffness actions in batches
local_f = lax.map(
    apply_element,
    (weights, local_u),
    batch_size=B_batch,
) |\label{algjax_galerkin_sliced:line:map}|              # local_f: (Mx*My, 8)
\end{minted}
\caption{Batched evaluation of the matrix-free Galerkin stiffness action using JAX.
The batch size $B_\mathrm{batch}$ controls the number of coarse elements processed simultaneously.}
\label{lst:galerkin_sliced}
\end{listing}

The resulting local force contributions retain the same shape as in the unsliced formulation and are assembled using the tensor manipulations starting at line~\ref{algjax_galerkin:line:assembly} of Listing~\ref{lst:galerkin_matrixfree2d}.
As in the fine-grid slicing approach, processing the batches sequentially does not eliminate GPU parallelism, since the contractions within each batch can still be evaluated in parallel.
Provided that the batch size is sufficiently large, this can limit intermediate memory requirements without substantially reducing GPU utilization.

The extension to three dimensions is straightforward.
Each hexahedral element has twenty-four displacement degrees of freedom, and an element at level $\ell$ contains $8^\ell$ original fine-grid elements.
Consequently, the Galerkin operator is represented by $8^\ell$ fixed matrices of size $24\times24$, and the local stiffness action becomes
\begin{align}
    \fvec_e^{(\ell)}
    &=
    \sum_{s=0}^{8^\ell-1}
    w_{e,s}G_s^{(\ell)}\uvec_e.
\end{align}

For a three-dimensional mesh containing $N_x^{(\ell)}\times N_y^{(\ell)}\times N_z^{(\ell)}$ coarse elements, evaluating the complete contraction involves $8^\ell \times N_x^{(\ell)} \times N_y^{(\ell)} \times  N_z^{(\ell)}$ subelement contributions.
Processing $B_\mathrm{batch}$ coarse elements at a time reduces the number of contributions evaluated per batch to $B_\mathrm{batch} \times 8^\ell$.
As in two dimensions, this limits the temporary storage associated with the contraction without changing the total computational work.

In the implementation, bounded batching is combined with spatial slicing of the coarse grid along the $x$-direction.
Each slab spans the full extent of the grid in the $y$- and $z$-directions, while the grouping of fine-grid weights, extraction of local displacements, Galerkin contraction, and assembly of nodal contributions are performed locally within the slab.
This avoids materializing the corresponding element-local intermediate arrays over the complete coarse grid.
The benefit becomes particularly relevant in three dimensions, where each local Galerkin operator contains $24\times24$ entries.
By evaluating the weighted contributions in bounded batches, we avoid materializing the spatially varying coarse-element stiffness matrices and limit the memory required for intermediate computations while preserving the Galerkin stiffness action.


\section{The \topax{} library}

\begin{figure*}[t]
  \centering
  \resizebox{\textwidth}{!}{%
    \begin{tikzpicture}[
	font=\sffamily\small,
	>={Stealth[length=2mm,width=1.35mm]},
	setup flow/.style={->, draw=topaxBlue, line width=0.8pt},
	jax flow/.style={->, draw=topaxGreen, line width=0.8pt},
	support/.style={
	-{Stealth[length=2.3mm,width=1.55mm]},
	draw=topaxBlue!85,
	line width=1pt,
	dash pattern=on 4pt off 2.5pt
	},
	stage/.style={
			draw=topaxInk,
			line width=0.65pt,
			rounded corners=1.5pt,
			fill=topaxPale,
			minimum height=11mm,
			minimum width=25mm,
			align=center,
			inner sep=4pt
		},
	major/.style={stage},
	setup/.style={stage, text width=36mm},
	setup container/.style={
			draw=topaxBlue!60,
			fill=topaxBluePale,
			line width=0.6pt,
			rounded corners=2pt
		},
	jax container/.style={
		draw=topaxGreen!75,
		fill=topaxGreenPale,
			line width=0.6pt,
			rounded corners=2pt
		},
	container title/.style={font=\sffamily\scriptsize\bfseries},
	transfer/.style={->, draw=topaxOrange, line width=0.82pt},
	level box/.style={
			draw=topaxInk,
			line width=0.65pt,
			rounded corners=1.5pt,
			fill=topaxPale,
			minimum height=10mm,
			align=center,
			inner sep=4pt
		},
	level/.style={level box, text width=61mm},
	coarse/.style={level, line width=0.9pt},
  edge label/.style={
    font=\sffamily\scriptsize,
    text=topaxMuted,
    inner sep=1.5pt
  },
	zoom link/.style={
	-{Stealth[length=2.4mm,width=1.6mm]},
	draw=topaxOrange,
	line width=1pt,
	dash pattern=on 4pt off 2.5pt
	},
	detail marker/.style={
			draw=topaxOrange,
			fill=topaxOrangePale,
			line width=0.9pt,
			dash pattern=on 3pt off 2pt,
			rounded corners=1pt
		},
	hierarchy container/.style={
			draw=topaxOrange,
			fill=topaxOrangePale,
			line width=1pt,
			dash pattern=on 4pt off 2.5pt,
			rounded corners=2pt
		}
	]

	\begin{scope}[scale=0.7, transform shape, local bounding box=flowPanel]
		\node[setup] (problem) at (-5.4,3.7) {
			\textbf{Problem definition}
		};
		\node[setup] (discretization) at (0,3.7) {
			\textbf{Structured discretization}
		};
		\node[setup] (sparse) at (5.4,3.7) {
			\textbf{Coarse-solver setup}
		};

		\draw[setup flow] (problem) -- (discretization);
		\draw[setup flow] (discretization) -- (sparse);

		\node[stage] (design) at (-6.50,0) {
			\textbf{Design variables}\\[-1pt]
			$\thetavec_t$
		};
		\node[stage] (filter) at (-2.70,0) {
			\textbf{Filter and project}
		};
		\node[major, minimum width=48mm, minimum height=18mm]
		(analysis) at (2.00,0) {};
		\node[detail marker, font=\sffamily\small\bfseries,
			inner xsep=2.5mm, inner ysep=0.8mm]
		(solveDetail) at ($(analysis.center)+(0,3.3mm)$) {Linear solve};
		\node[font=\sffamily\small\bfseries]
		at ($(analysis.center)+(0,-3.4mm)$) {and sensitivities};

		\node[stage] (mma) at (6.55,0) {
			\textbf{Design update}\\[-1pt]
			$\thetavec_{t+1}$
		};

		\draw[jax flow] (design) -- (filter);
		\draw[jax flow] (filter) -- (analysis);
		\draw[jax flow] (analysis) -- (mma);
		\draw[jax flow, rounded corners=3pt]
		(mma.south) -- ++(0,-1.05) -| (design.south);

		\draw[support] (discretization.south) -- (filter.north);
		\draw[support] ($(discretization.south)+(0.45,0)$) -- (analysis.north);
		\draw[support] (sparse.south) -- (analysis.north east);

		\coordinate (setupSW) at
		($(design.west |- problem.south)+(-4mm,-4mm)$);
		\coordinate (setupNE) at
		($(mma.east |- problem.north)+(4mm,8mm)$);
		\coordinate (jaxSW) at ($(design.south west)+(-4mm,-12mm)$);
		\coordinate (jaxNE) at
		($(mma.east |- analysis.north)+(4mm,8mm)$);

		\begin{scope}[on background layer]
			\path[setup container] (setupSW) rectangle (setupNE);
			\path[jax container] (jaxSW) rectangle (jaxNE);
		\end{scope}

		\node[container title, text=topaxBlue, anchor=north east]
		at ($(setupNE)+(-2mm,-2mm)$) {One-time setup};
	\node[container title, text=topaxGreen, anchor=north east]
		at ($(jaxNE)+(-2mm,-2mm)$) {Jittable pure JAX function};
	\end{scope}

	\begin{scope}[
			xshift=9.2cm,
			yshift=1.6cm,
			scale=0.6,
			transform shape,
			local bounding box=hierarchyPanel
		]
		\node[level] (levelzero) at (0,1.15) {
			\textbf{Fine level $\ell{=}0$}\\[-1pt]
			\scriptsize matrix-free action using $\stiffness_e{=}w_eJ$
		};
		\node[level] (implicit) at (0,-0.15) {
			\textbf{Implicit Galerkin levels $\ell {\in} \{1,\ldots,{L_\mathrm{cross}}{-}1 \}$}\\[-1pt]
			\scriptsize matrix-free operator using
			$G_s=Q_s^{\mathsf T} \elementstiffness Q_s$
		};
		\node[level] (heterogeneous) at (0,-1.45) {
			\textbf{Materialized levels $\ell {\in} \{L_{\mathrm{cross}},\ldots,L-2\}$}\\[-1pt]
			\scriptsize one dense local stiffness matrix per element
		};
		\node[coarse] (coarsest) at (0,-2.75) {
			\textbf{Coarsest level $\ell{=}{L-1}$}\\[-1pt]
			\scriptsize assembled lower-triangular CSR
			$\rightarrow$ sparse Cholesky
		};

		\draw[transfer] ($(levelzero.west)+(-0.18,0)$)
		to[bend right=18] node[edge label, left] {$P^{\mathsf T}$}
		($(implicit.west)+(-0.18,0)$);
		\draw[transfer] ($(implicit.west)+(-0.18,0)$)
		to[bend right=18] node[edge label, left] {$P^{\mathsf T}$}
		($(heterogeneous.west)+(-0.18,0)$);
		\draw[transfer] ($(heterogeneous.west)+(-0.18,0)$)
		to[bend right=18] node[edge label, left] {$P^{\mathsf T}$}
		($(coarsest.west)+(-0.18,0)$);

		\draw[transfer] ($(coarsest.east)+(0.18,0)$)
		to[bend right=18] node[edge label, right] {$P$}
		($(heterogeneous.east)+(0.18,0)$);
		\draw[transfer] ($(heterogeneous.east)+(0.18,0)$)
		to[bend right=18] node[edge label, right] {$P$}
		($(implicit.east)+(0.18,0)$);
		\draw[transfer] ($(implicit.east)+(0.18,0)$)
		to[bend right=18] node[edge label, right] {$P$}
		($(levelzero.east)+(0.18,0)$);

		\begin{scope}[on background layer]
			\node[hierarchy container, fit=(levelzero)(coarsest),
				inner xsep=7mm, inner ysep=5mm] (hierarchyFrame) {};
		\end{scope}
		\node[container title, text=topaxOrange, anchor=north east]
		at ($(hierarchyFrame.north east)+(-2mm,-2mm)$)
		{Linear-solve hierarchy};
	\end{scope}

	\draw[zoom link]
	(hierarchyFrame.south west)
	.. controls +(-1.5cm,-0.8cm) and +(1.0cm,-1.8cm) ..
	(solveDetail.south east);

\end{tikzpicture}%
  }
  \caption{Overview of the \topax{} topology-optimization workflow.
  Static discretization data and sparse-solver resources are constructed once (blue box), while filtering and projection, the linear solve, sensitivity evaluation, and the design update form the compiled design iteration (green box).}
  \label{fig:topax_overview}
\end{figure*}
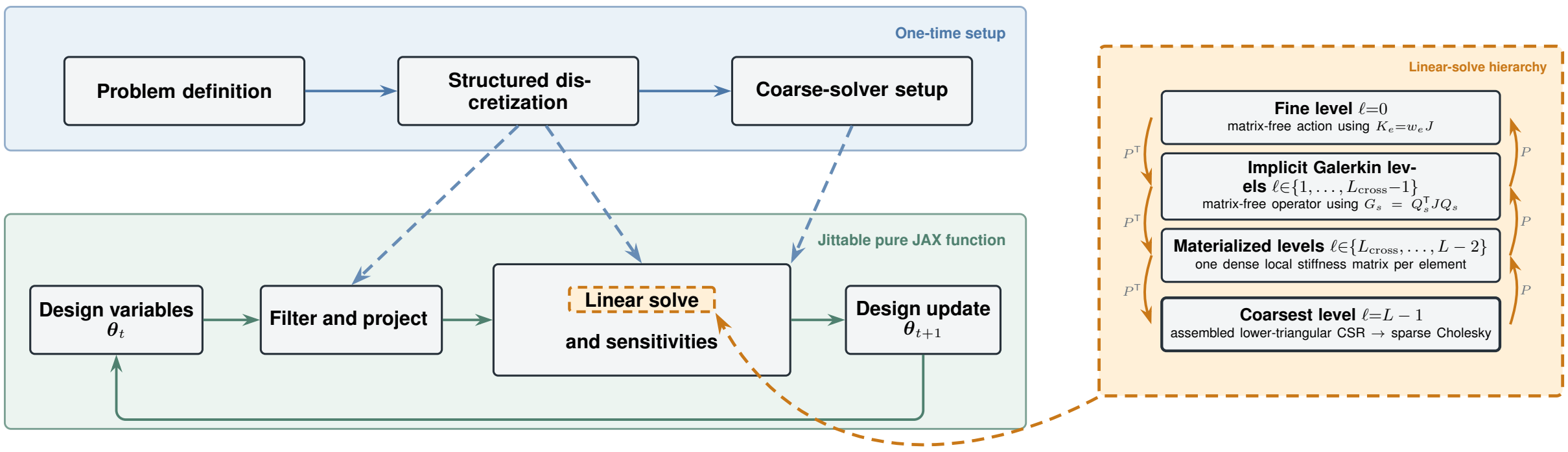

We implement the proposed matrix-free and Galerkin operators in \topax{}, a modular topology-optimization library developed using JAX~\citep{jax2018github}.
The library provides a flexible framework for density-based topology optimization, allowing different numerical methods to be combined without modifying the underlying finite-element implementation.
It supports Cartesian quadrilateral and hexahedral meshes in two and three dimensions and can execute on both CPUs and GPUs.
Figure~\ref{fig:topax_overview} illustrates the overall architecture of the library, including the topology-optimization workflow and the multigrid hierarchy used for finite-element analysis.
The implementation is released as open-source software and is publicly available at \href{https://github.com/pedrozudo/topax}{https://github.com/pedrozudo/topax}.

\subsection{Library architecture}

\topax{} separates the main components of the topology-optimization problem into interchangeable modules.
The optimization algorithm can be selected independently of the finite-element solver, while the filtering and projection operations can be configured according to the problem requirements.
In particular, the library supports both convolution-based density filtering and Helmholtz-type filtering, as well as different projection functions for mapping the filtered design variables to physical densities.

The linear solver is similarly modular.
The finite-element equilibrium equations are solved using multigrid preconditioned conjugate gradients, with configurable smoothing procedures and a replaceable sparse direct solver for the coarsest level.
This separation allows individual numerical components to be replaced or extended without changing the overall optimization workflow.

The implementation relies on JAX for tensor operations, automatic differentiation, and just-in-time compilation.
By expressing the numerical computations through standard JAX primitives, the same implementation can target different hardware architectures, including CPUs and GPUs.
The structured discretization and tensor-based operators are formulated for two as well as three spatial dimensions, allowing the main numerical algorithms to share the same implementation.

\subsection{Construction of the multigrid hierarchy}

The multigrid hierarchy combines different representations of the stiffness operator depending on the grid level.
The finest level uses the sliced convolution-based matrix-free operator developed in Section~\ref{sec:free_slicing}.
On the subsequent levels, the Galerkin stiffness action is evaluated using the implicit tensor contractions introduced in Section~\ref{sec:galerkin_slicing}.
Although the implicit Galerkin representation avoids materializing a separate stiffness matrix for each coarse element, the number of precomputed operators increases with the coarsening level.
In two dimensions, each element at level $\ell$ requires $4^\ell$ operators, while in three dimensions the corresponding number is $8^\ell$.
Consequently, evaluating the weighted contributions becomes increasingly expensive on deeper levels of the hierarchy.
To address this, \topax{ }introduces a configurable crossover level $L_{\mathrm{cross}}$.
For levels $\ell\in\{1,\ldots,L_{\mathrm{cross}}-1\}$, the Galerkin action is evaluated implicitly using the original fine-grid stiffness interpolation coefficients and the precomputed operators $G_s^{(\ell)}$ (\cf Section~\ref{sec:galerkin_slicing}).
At level $L_\mathrm{cross}$, we explicitly construct the individual local stiffness matrices per element using Equation~\ref{eq:galerkin_weighted_operators}.
That is, we construct a separate stiffness matrix for each element once the problem has been sufficiently coarsened.

Once these matrices have been constructed, the remaining levels are obtained through recursive element-wise Galerkin projection.
Each parent element combines the stiffness matrices of its four children in two dimensions, or eight children in three dimensions.
In contrast to the implicit formulation, this representation stores one stiffness matrix per element but avoids repeatedly summing contributions from all the original fine-grid subelements.
The materialized element matrices are applied through local matrix--vector products followed by assembly of the resulting nodal forces.
Thus, the intermediate levels continue to avoid assembling a global sparse stiffness matrix.
The crossover level therefore controls the balance between the computational cost of implicit Galerkin contractions and the memory required to store individual stiffness matrices for each element.
It can be selected according to the problem size and available hardware memory.

The global stiffness matrix is stored explicitly only at the coarsest level, where it is assembled in compressed sparse row (CSR) format.
The resulting linear system is solved using sparse Cholesky factorization, with CHOLMOD~\citep{chen2008algorithm} on CPUs and cuDSS\footnote{\href{https://developer.nvidia.com/cudss}{https://developer.nvidia.com/cudss}} on NVIDIA GPUs.
The sparsity pattern and symbolic factorization can be established during initialization, while the numerical factorization is updated as the material distribution changes throughout the optimization.

\subsection{Optimization workflow}

The topology-optimization workflow is illustrated in Figure~\ref{fig:topax_overview}.
Starting from the design variables, the selected filtering method produces a filtered density field, which is subsequently transformed into the physical densities through the chosen projection function.
These densities determine the element stiffness interpolation coefficients according to the SIMP model.
For each material distribution, the library updates the stiffness coefficients and the corresponding multigrid data.
The finite-element equilibrium equations are then solved using the MGPCG method introduced in Section~\ref{sec:mgcg}.

After solving the equilibrium equations, the objective and constraint functions are evaluated together with their sensitivities.
JAX provides automatic differentiation of the numerical operations, allowing the required derivatives to be obtained without manually implementing the corresponding sensitivity expressions.
Note that the differentiation through the linear solve is implemented via the adjoint method and does not unroll the conjugate gradient loop.
The resulting objective and constraint sensitivities are passed to the selected optimization algorithm, which updates the design variables.
The process is repeated until the prescribed convergence criterion is reached.

As illustrated in Figure~\ref{fig:topax_overview}, the library distinguishes between operations that depend only on the problem definition and those that must be repeated during optimization.
Concretely, the structured discretization and sparse-solver resources are initialized separately from the repeated optimization computations.
The mesh structure and the sparsity pattern of the coarsest stiffness matrix remain fixed throughout the optimization, while their material-dependent numerical values are updated as the design evolves.
This allows filtering, projection, finite-element analysis, sensitivity evaluation, and design updates to be integrated into a compiled JAX workflow while keeping the individual numerical components configurable.


\section{Related work}

\subsection{Large-scale topology optimization}

\citet{amir2014multigrid} introduced multigrid preconditioning to topology optimization, establishing an approach that has since become fundamental to many large-scale implementations.
Although their experiments were limited to relatively modest problem sizes, subsequent work demonstrated the scalability of multigrid-based methods.
In particular, \citet{aage2015topology} developed a massively parallel topology optimization framework based on PETSc~\citep{balay1997efficient}, laying the groundwork for gigavoxel-scale topology optimization using distributed-memory computing~\citep{aage2017giga,baandrup2020closing}.

Matrix-free finite-element formulations provide another means of reducing memory consumption in large-scale topology optimization by avoiding the assembly and storage of the global stiffness matrix.
Recently, \citet{wang2025efficient} combined matrix-free finite-element analysis with geometric multigrid preconditioning in a MATLAB implementation capable of handling problems with up to 128 million hexahedral elements on a workstation with 64 GB of memory.
Their approach evaluates element-based matrix operations in memory-efficient chunks, while employing non-dyadic coarsening to reduce the memory requirements of the multigrid hierarchy.
In contrast, our formulation exploits structured mesh connectivity through convolutions and tensor transformations, and retains a dyadic Galerkin hierarchy with implicit coarse-grid operator evaluation.

\subsection{GPU acceleration}

GPU acceleration has been widely investigated as a means of improving the computational efficiency of topology optimization.
\citet{schmidt20112589} demonstrated an early implementation of topology optimization on graphics hardware.
\citet{dick2011real} developed a matrix-free geometric multigrid solver for elasticity simulation using hexahedral finite elements, exploiting regular mesh connectivity through a specialized CUDA implementation.
Building on this approach, \citet{wu2015system} developed a GPU-accelerated system for high-resolution topology optimization, reducing memory requirements through on-the-fly construction of finest-grid stencils and memory-efficient multigrid techniques.
\citet{martinez2017efficient} further investigated GPU-based topology optimization using multilevel granularity, with specialized CUDA kernels for finite-element analysis, sensitivity calculations, filtering, and design updates.
\citet{ratnakar2021gpu} investigated a GPU-based matrix-free conjugate gradient solver with customized nodal connectivity storage.
More recently, \citet{yang2026simp} developed fused gather--matrix-multiplication--scatter kernels for three-dimensional SIMP topology optimization, reducing intermediate memory traffic by combining element-level operations into a single CUDA kernel.
Unlike these specialized GPU implementations, our formulation expresses the stiffness action through convolutions and structured tensor transformations that can leverage hardware-optimized numerical kernels without requiring custom CUDA kernels.

The representation of Galerkin coarse-grid operators introduces additional computational and memory requirements in GPU-based multigrid solvers.
Although \citet{dick2011real}, \citet{wu2015system}, and \citet{martinez2017efficient} employ Galerkin multigrid, their implementations explicitly construct and store coarse-grid stencils or coefficient matrices.
In particular, \citet{martinez2017efficient} identify the memory requirements and global-memory accesses associated with these coefficients as limitations of their implementation.
More recently, \citet{yang2026galerkin} developed a Galerkin multigrid solver for three-dimensional topology optimization that combines matrix-free fine-grid evaluation with explicitly constructed coarse-grid operators.
Their first coarse level is obtained through local Galerkin aggregation, while subsequent levels are constructed using sparse Galerkin products.

Our approach differs in how these Galerkin operators are represented and evaluated.
Rather than using local Galerkin projections solely to construct explicitly stored coarse-element matrices or nodal stencils, we retain a factorized representation consisting of fine-grid stiffness interpolation coefficients and precomputed local operators.
This representation is evaluated directly during the multigrid V-cycle through batched tensor contractions, without materializing the coarse-element stiffness matrices on the selected implicit levels.
At sufficiently coarse levels, the implementation transitions to explicitly stored operators, allowing a configurable trade-off between memory consumption and the cost of repeated operator evaluation.

\subsection{High-level and differentiable programming}

High-level programming frameworks have increasingly been used to develop hardware-accelerated topology optimization methods without requiring specialized low-level implementations.
\citet{traff2023simple} presented GPU-accelerated topology optimization implementations using OpenMP~\citep{dagum1998openmp} and Futhark~\citep{henriksen2017futhark}, demonstrating that high-level parallel programming frameworks can achieve competitive performance without specialized CUDA kernels.
Their work also highlights challenges associated with memory consumption, particularly in the Futhark implementation, where compiler-generated intermediate arrays contribute to substantially higher memory usage.
More recently, \citet{nobari2026pgdtoscalablealternativemma} introduced pyFANTOM, a modular Python framework for topology optimization with separate CPU and GPU backends. Its GPU implementation employs CuPy~\citep{nishino2017cupy} and specialized CUDA kernels for matrix-free finite-element analysis and geometric multigrid, rather than high-level tensor operations.

Automatic differentiation provides another motivation for adopting high-level numerical frameworks, as it simplifies sensitivity analysis and facilitates the development of flexible optimization methods.
\citet{norgaard2017applications} investigated the application of automatic differentiation to topology optimization.
Subsequently, \citet{chandrasekhar2021auto} introduced AuTO, an educational JAX-based framework demonstrating automatic differentiation for structural optimization using dense finite-element matrix assembly and direct linear solvers.
Other Python-based frameworks wrap existing numerical solvers to support differentiable finite-element analysis.
\citet{xue2023jax} developed JAX-FEM, which integrates automatic differentiation with sparse matrix assembly and existing linear solver backends.
Similarly, \citet{jia2024fenitop} introduced FEniTop, a Python-based SIMP topology optimization framework that wraps FEniCSx~\citep{BarattaEtal2023} and PETSc for finite-element assembly and linear system solution, using symbolic differentiation for sensitivity analysis.
More recent work has explored automatic differentiation for structural shape optimization~\citep{wu2023framework}, level-set topology optimization~\citep{neofytou2024level}, neural topology optimization~\citep{sanu2026leveraging}, and multiphysics optimization~\citep{padhy2026toflux}.

While these approaches demonstrate the benefits of high-level programming and automatic differentiation, the efficiency of the underlying numerical operations remains critical for large-scale topology optimization.
\citet{traff2023simple} report increased memory consumption in their Futhark implementation due to compiler-generated intermediate arrays.
Among the differentiable frameworks, AuTO assembles dense stiffness matrices, while JAX-FEM and FEniTop rely on sparse matrix assembly.
Although sparse assembly substantially reduces storage compared with dense matrices, these approaches retain the memory overhead of explicitly assembled stiffness operators, limiting their applicability to very large three-dimensional problems.
In contrast, \topax{} expresses the fine-grid stiffness action through convolutions and structured tensor transformations, and evaluates Galerkin coarse-grid operators using batched tensor contractions.
Slicing and batching limit intermediate memory requirements, while JAX provides access to hardware-optimized computational kernels, just-in-time compilation, and automatic differentiation.
The resulting formulation combines the flexibility of a high-level differentiable programming environment with high computational performance on CPUs and GPUs.


\section{Two-dimensional validation experiments}

This section presents two two-dimensional topology-optimization examples from the literature to validate the implementation and assess the computational scalability of \topax{} in two dimensions.
All two-dimensional experiments were performed on a workstation equipped with an NVIDIA GeForce RTX~4090 GPU with $24\,\mathrm{GiB}$ of device memory.

\subsection{Density filter with linearly decaying weighting}
\label{sec:sigmund_density_filter}

We first consider the MBB-beam example studied by \citet{sigmund2007morphology}.
The original problem is defined on a rectangular design domain with dimensions $120\times 40$ and uniform thickness $t=1$, and is discretized using $120\times40$ elements.
Plane-stress conditions are assumed.
The horizontal displacement is constrained along the left boundary, while the vertical displacement is constrained at the lower-right corner.
A downward point load of magnitude $1$ is applied at the upper-left corner.
The prescribed volume fraction is $0.5$, the SIMP penalty exponent is $p=3$, the Young's modulus is $E=1$, the Poisson's ratio is $\nu=0.3$, and the minimum relative stiffness is $10^{-9}$.
In addition to the original discretization, we consider the geometrically scaled meshes $360\times120$, $750\times250$, $1500\times500$, $3000\times1000$, and $5400\times1800$.
These larger cases are intended to evaluate computational scaling and final-design quality; they do not constitute a mesh-convergence study of a fixed physical domain.
The element dimensions are fixed at $1\times1$ for every mesh.
Consequently, increasing the number of elements enlarges the physical design domain while preserving its $3{:}1$ aspect ratio.

The design variables are regularized using the standard density filter with linearly decaying weighting introduced by \citet{bourdin2001filters} and considered by \citet{sigmund2007morphology}.
On the structured grids used by \topax{}, this filter maps directly onto the JAX \texttt{lax.conv\_general\_dilated} primitive.
The filter radius is fixed at $R=3.5$ element widths for all meshes.
More precisely, the filter includes elements whose centroids lie within a Euclidean distance of $3.5$ times the element width from the centroid of the element being filtered.
Their contributions decrease linearly with distance and vanish at the filter radius.
For the unit-sized square elements considered here, this produces a $7\times7$ bounding stencil with circular support.
The same stencil is used for every mesh.
Therefore, these experiments at different mesh sizes represent geometrically scaled problems rather than successive mesh refinements of a fixed physical domain.
Furthermore, no Heaviside projection or other thresholding operation is applied.
Consequently, the filtered field directly represents the physical density used in the SIMP interpolation, finite-element analysis, volume constraint, and visualization of the optimized designs.

The optimization problems are solved using the method of moving asymptotes (MMA)~\citep{svanberg1987method}.
The first three meshes use two multigrid levels, while the three largest meshes use three levels.
In both configurations, the fine-grid stiffness operator is applied matrix-free using the slicing strategy described in Section~\ref{sec:slicing_matrix_free}, which limits the memory required by intermediate tensors.
In the two-level configuration, the second level is the coarsest level; its stiffness matrix is assembled and the corresponding system is solved directly.
In the three-level configuration, the intermediate Galerkin operator is evaluated without assembling or storing its global stiffness matrix.
Its element-wise contributions are processed in bounded batches, as described in Section~\ref{sec:sliced_galerkin}.
Only the coarsest stiffness matrix is assembled, and its system is solved directly.

Optimization terminates when either of two stopping criteria is satisfied.
The first criterion is a maximum change in the design variables of at most $0.01$.
The second criterion detects a plateau in the compliance and is activated only after at least 500 design iterations.
At iteration $k{\geq}500$, the compliance is considered to have plateaued if
$
    \nicefrac{
        |C^{(k)}-C^{(k-100)}|
    }{
        |C^{(k)}|
    }
    \leq 10^{-5}
$.
This plateau criterion compares the current compliance with that obtained 100 iterations earlier and is evaluated at every iteration after its activation.
A maximum of $10\,000$ design iterations is imposed.

\begin{figure*}[h]
    \centering
    \includegraphics[width=0.95\textwidth]{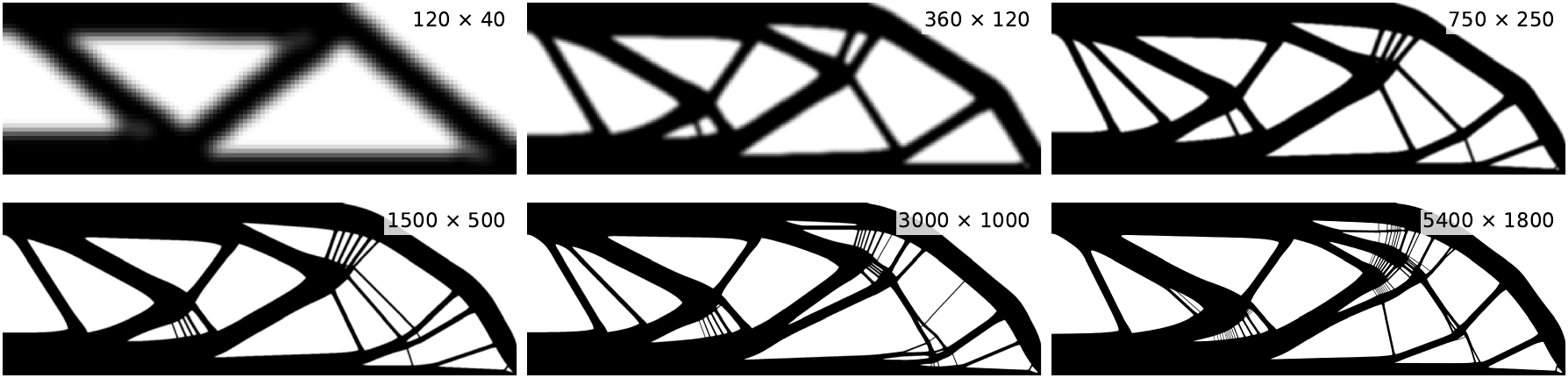}
    \caption{
        Physical-density fields obtained using the density filter with
        linearly decaying weighting. Mesh size increases from left to right
        and from top to bottom. No projection is applied.
    }
    \label{fig:sigmund_designs}
\end{figure*}

Figure~\ref{fig:sigmund_designs} shows the physical density fields at termination.
The $120\times40$ case is consistent with the reference design reported by \citet{sigmund2007morphology}.
The layouts obtained on the larger meshes differ from the coarse-mesh result and contain more numerous and finer structural members.
Because the filter radius remains fixed in element units as the mesh dimensions increase, the imposed length scale decreases relative to the dimensions of the design domain.
These cases therefore do not constitute successive mesh refinements of a fixed physical problem.

For the original $120\times40$ problem, \topax{} obtains a compliance of $218.1546$, compared with the value $218.3$ reported by \citet{sigmund2007morphology}.
This corresponds to a relative difference of approximately $-0.067\%$.
The \topax{} run satisfies the design-change criterion after 452 iterations.
By comparison, the reference result reported by \citeauthor{sigmund2007morphology} was obtained after 1000 iterations.

The degree to which the physical density field contains intermediate densities is quantified using the measure of non-discreteness ($M_{\mathrm{nd}}$) introduced by \citet{sigmund2007morphology}:
\begin{align}
    M_{\mathrm{nd}}
    =
    \frac{100}{N_e}
    \sum_{e=0}^{N_e-1}
    4\rho_e
    \left(1-\rho_e\right).
\end{align}
A physical density of either zero or one makes no contribution to $M_{\mathrm{nd}}$, whereas a physical density of $0.5$ gives the maximum contribution.
Consequently, $M_{\mathrm{nd}}=0\%$ represents a completely discrete design, while larger values indicate a greater presence of intermediate densities.
For the original problem, \topax{} obtains $M_{\mathrm{nd}}=23.933\%$, compared with the value $24.02\%$ reported by \citeauthor{sigmund2007morphology}.
This corresponds to a relative difference of approximately $-0.362\%$.

Table~\ref{tab:sigmund} summarizes the computational results.
The reported average time per design iteration excludes compilation and the untimed warm-up iteration.
The mean MGPCG iteration count refers to the stiffness solve performed during each design iteration.
Peak GPU-memory usage was determined by sampling the device memory at $0.2\,\mathrm{s}$ intervals throughout each run and subtracting the baseline measured immediately before the experiment.
JAX GPU-memory preallocation was disabled, such that the reported values reflect the incremental high-water mark of allocations made during the experiment.
The complete measurement procedure and experimental configuration are documented in the experiment scripts distributed with \topax{}.

\begin{table*}[h]
\centering
\scriptsize
\setlength{\tabcolsep}{4pt}
\caption{Optimization and computational results for the geometrically scaled MBB-beam problems.}
\label{tab:sigmund}
\begin{tabular}{rrrrrrrrr}
\toprule
mesh &
elements &
levels &
design iter. &
\shortstack{avg. time per iter. [s]} &
\shortstack{avg. MGPCG iter.} &
\shortstack{peak mem. [GiB]} &
compliance &
\shortstack{non-discreteness \mbox{[\%]}} \\
\midrule
120$\times$40 & 4,800 & 2 & 452 & 0.010 & 14.42 & 0.611 & 218.1546 & 23.933 \\
360$\times$120 & 43,200 & 2 & 3,581 & 0.022 & 13.07 & 0.666 & 196.7388 & 13.310 \\
750$\times$250 & 187,500 & 2 & 4,636 & 0.050 & 12.40 & 0.875 & 192.5300 & 8.262 \\
1500$\times$500 & 750,000 & 3 & 7,850 & 0.208 & 16.32 & 1.053 & 191.3681 & 4.622 \\
3000$\times$1000 & 3,000,000 & 3 & 10,000 & 0.549 & 17.93 & 2.361 & 192.5512 & 2.713 \\
5400$\times$1800 & 9,720,000 & 3 & 10,000 & 1.416 & 19.26 & 6.395 & 194.5227 & 1.661 \\
\bottomrule
\end{tabular}
\end{table*}

The mean design-iteration time increases from $0.010\,\mathrm{s}$ for
$4\,800$ elements to $1.416\,\mathrm{s}$ for $9.72$ million elements.
Over the same range, the baseline-corrected peak GPU-memory consumption increases
from $0.611\,\mathrm{GiB}$ to $6.395\,\mathrm{GiB}$.
The largest problem therefore fits comfortably on the $24\,\mathrm{GiB}$ NVIDIA GeForce RTX~4090 used for these experiments.
The mean number of stiffness-MGPCG iterations remains moderate, increasing from approximately $14.42$ to $19.26$.

The non-discreteness decreases from $23.933\%$ for the original problem to $1.661\%$ for the largest problem.
The increasingly sharp physical-density fields shown in Figure~\ref{fig:sigmund_designs} can be explained by keeping the filter stencil fixed in element units while increasing the mesh dimensions.
Consequently, the filter radius and the associated transition regions become progressively smaller relative to the complete design domain.
The decrease in $M_{\mathrm{nd}}$ therefore reflects the changing ratio between the filter radius and the design-domain dimensions, rather than mesh convergence for a fixed physical problem.

The $3000\times1000$ and $5400\times1800$ cases both reach the imposed limit of $10\,000$ design iterations without satisfying either stopping criterion.
Their reported compliance values therefore correspond to the final optimization iterates, and the resulting topologies should not be interpreted as converged according to the prescribed criteria.
Nevertheless, the experiments demonstrate that \topax{} can optimize a two-dimensional problem containing nearly ten million elements on consumer GPU hardware while maintaining moderate MGPCG iteration counts and a memory footprint below the available device capacity.

\subsection{Helmholtz-type density filter}
\label{sec:lazarov_density_filter}

We next consider the cantilever example used by \citet{lazarov2011filters}.
The design domain has dimensions $4\times1$ and a thickness of $t=0.01$, and plane-stress conditions are assumed.
A downward point load of magnitude $0.2$ is applied at the midpoint of the right boundary, while the left boundary is fully fixed.
The prescribed volume fraction is $0.5$, the SIMP penalty exponent is $p=3$, the Young's modulus is $E=1$, the Poisson's ratio is $\nu=0.2$, and the minimum relative stiffness is $10^{-9}$.

The design variables are regularized using the Helmholtz-type density filter introduced by \citet{lazarov2011filters}.
The filter radius is fixed at $r=0.02$ in physical dimensions for every mesh.
Consequently, in contrast to the geometrically scaled MBB beam example (\cf Section~\ref{sec:sigmund_density_filter}), these cases represent successive refinements of the same physical problem.
No projection or thresholding operation is applied, and the filtered field is used directly as the physical density in the SIMP interpolation, volume constraint, and visualization.

The three meshes considered by \citeauthor{lazarov2011filters} contain $400\times100$, $800\times200$, and $1200\times300$ elements.
We additionally consider meshes containing $1600\times400$, $3200\times800$, and $6400\times1600$ elements, with the largest problem containing $10.24$ million elements.
The design variables are updated using the optimality-criteria method (OC)~\citep{bendsoe2004topology}, with a move limit of $0.2$.
The first three meshes use two multigrid levels, while the three largest meshes use three levels.
The same matrix-free slicing, implicit Galerkin batching, and direct coarse-grid solution strategies described in the preceding subsection are employed.
Both the stiffness systems and the forward and adjoint Helmholtz systems are solved using MGPCG.
The complete set of solves is included in the reported design-iteration time, while the mean MGPCG iteration count reported in Table~\ref{tab:lazarov} refers specifically to the stiffness solve.
We use the stopping criteria and the GPU, timing, and memory-measurement procedures described in the preceding subsection.

\begin{figure*}[h]
    \centering
    \includegraphics[width=\textwidth]
        {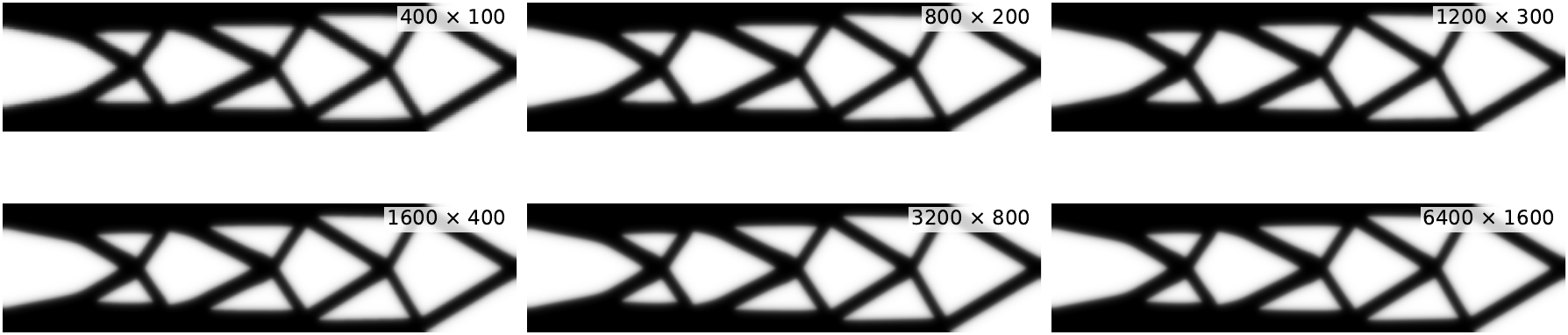}
    \caption{
        Physical-density fields obtained using Helmholtz density
        filter with $r=0.02$. Mesh resolution increases from left to right
        and from top to bottom. No projection is applied.
    }
    \label{fig:lazarov-designs}
\end{figure*}

Figure~\ref{fig:lazarov-designs} shows the physical-density fields obtained at the six mesh resolutions.
The optimized topologies have the same overall layout, with the same arrangement of structural members and openings across the different mesh resolutions.
The limited variation between the resulting designs indicates effective control of mesh dependence for this example, as expected when using a Helmholtz-type density filter with a constant filter radius $r$.
Since no projection is applied, smooth transitions between solid and void regions remain visible at every resolution.

To match the compliance convention used by \citet{lazarov2011filters}, the reported comparison value is obtained by multiplying the raw two-dimensional compliance by the thickness $t=0.01$.
For the three meshes reported in the original paper, \topax{} obtains compliance values of $16.5620$, $16.5369$, and $16.5329$, compared with the respective reference values $16.6556$, $16.6690$, and $16.6785$.
The corresponding relative differences are approximately $-0.56\%$, $-0.79\%$, and $-0.87\%$.

Table~\ref{tab:lazarov} summarizes the computational results.
The timing and memory quantities follow the definitions introduced in Section~\ref{sec:sigmund_density_filter}.
The measure of non-discreteness remains close to $26\%$ across all six resolutions.
Unlike the decreasing values observed for the geometrically scaled MBB beam problems, this nearly constant value is expected because both the physical domain and the physical filter radius are held fixed during mesh refinement.

\begin{table*}[h]
\centering
\scriptsize
\setlength{\tabcolsep}{4pt}
\caption{Optimization and computational results for the mesh-refined cantilever problems.}
\label{tab:lazarov}
\begin{tabular}{rrrrrrrrr}
\toprule
mesh &
elements &
levels &
design iter. &
\shortstack{avg. time per iter. [s]} &
\shortstack{avg. MGPCG iter.} &
\shortstack{peak mem. [GiB]} &
compliance &
\shortstack{non-discreteness \mbox{[\%]}} \\
\midrule
400$\times$100 & 40,000 & 2 & 1,305 & 0.037 & 10.19 & 0.822 & 16.5620 & 26.418 \\
800$\times$200 & 160,000 & 2 & 2,897 & 0.075 & 9.17 & 1.002 & 16.5369 & 26.121 \\
1200$\times$300 & 360,000 & 2 & 5,773 & 0.116 & 8.45 & 1.344 & 16.5329 & 26.048 \\
1600$\times$400 & 640,000 & 3 & 5,927 & 0.219 & 8.72 & 1.199 & 16.5382 & 26.026 \\
3200$\times$800 & 2,560,000 & 3 & 6,819 & 0.445 & 8.09 & 2.445 & 16.5531 & 25.999 \\
6400$\times$1600 & 10,240,000 & 3 & 6,548 & 1.136 & 7.43 & 7.398 & 16.5714 & 25.996 \\
\bottomrule
\end{tabular}
\end{table*}

All six optimization problems satisfy one of the prescribed stopping criteria.
The $400\times100$ and $800\times200$ cases terminate according to the design-change criterion, while the remaining cases terminate according to the compliance-plateau criterion.

The mean design-iteration time increases from $0.037\,\mathrm{s}$ for $40\,000$ elements to $1.136\,\mathrm{s}$ for $10.24$ million elements.
Over the same range, the baseline-corrected peak GPU-memory consumption increases from $0.822\,\mathrm{GiB}$ to $7.398\,\mathrm{GiB}$.
The mean number of stiffness-MGPCG iterations remains between approximately $7.4$ and $10.2$, with no systematic increase under mesh refinement.
The largest case therefore demonstrates the application of \topax{} to a topology-optimization problem with more than ten million finite elements on commodity hardware (NVIDIA GeForce RTX~4090), while retaining the same overall structural layout obtained at the lower mesh resolutions using the same physical filter radius.

\section{Large-scale three-dimensional numerical experiments}\label{sec:large_scale_three_dimensional}

\subsection{Large-scale Galerkin multigrid on GPUs}
\label{sec:large_scale_galerkin_gpu}

We compare \topax{} with two released GPU implementations that construct their multigrid hierarchies using Galerkin coarse-grid operators: Futtop\footnote{\href{https://github.com/topopt/futtop}{https://github.com/topopt/futtop}}, the Futhark implementation of \citet{traff2023simple}, and pyFANTOM\footnote{\href{https://github.com/ahnobari/pyFANTOM}{https://github.com/ahnobari/pyFANTOM}}~\citep{nobari2026pgdtoscalablealternativemma}.

The benchmark is the three-dimensional linear-elastic cantilever considered by \citet{traff2023simple}.
The rectangular cuboid design domain has fixed dimensions $2\times1\times1$ and is discretized using successively finer meshes.
All displacement components are constrained over two strips on one end face, and a distributed load with a total magnitude of one acts in the negative $z$-direction over a semicircular patch near the opposite end.
The loading and boundary conditions follow the original benchmark; their complete definitions are available in the cited paper and in the source code accompanying this work.

We consider 11 meshes ranging from $64\times32\times32$ to $640\times320\times320$ trilinear hexahedral elements, corresponding to $65\,536$ through $65\,536\,000$ elements.
The prescribed volume fraction is $0.1$.
A SIMP penalty exponent of $p=3$, a Young's modulus of $E=1$, a Poisson's ratio of $\nu=0.3$, and a minimum relative stiffness of $10^{-6}$ are used.
The design variables are regularized using a density filter with linearly decaying weights and a radius of $1.5$ element widths.
No projection or thresholding operation is applied.
All three implementations update the design using the optimality-criteria method with a move limit of $0.2$.
Each problem is run for 20 design iterations.
This follows the experimental setup of \citet{traff2023simple}.

\subsubsection*{Runtime and memory scaling}

\begin{figure*}[h]
    \centering
    \includegraphics[width=0.32\textwidth]
        {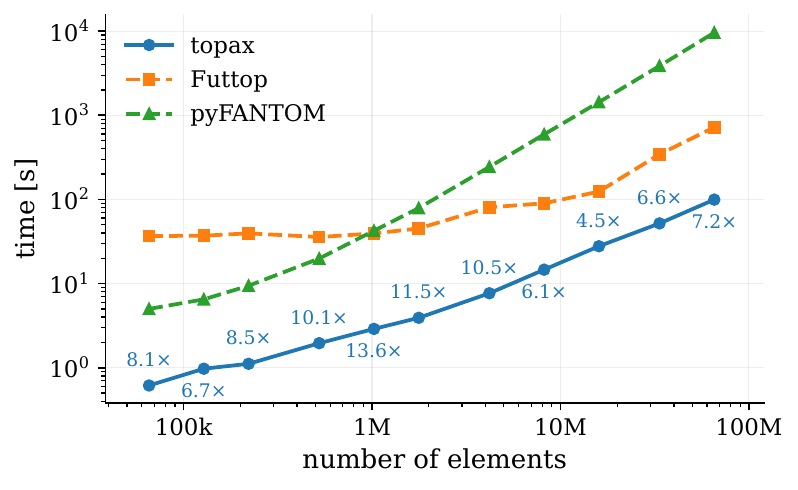}
    \hfill
    \includegraphics[width=0.32\textwidth]
        {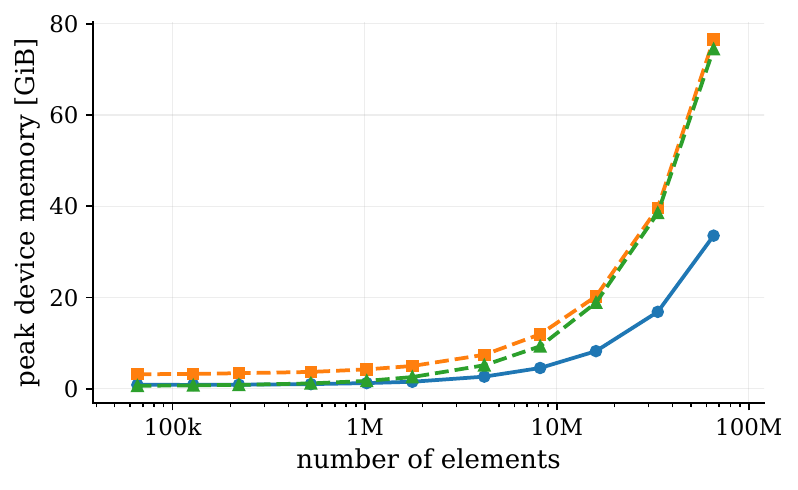}
    \hfill
    \includegraphics[width=0.32\textwidth]
        {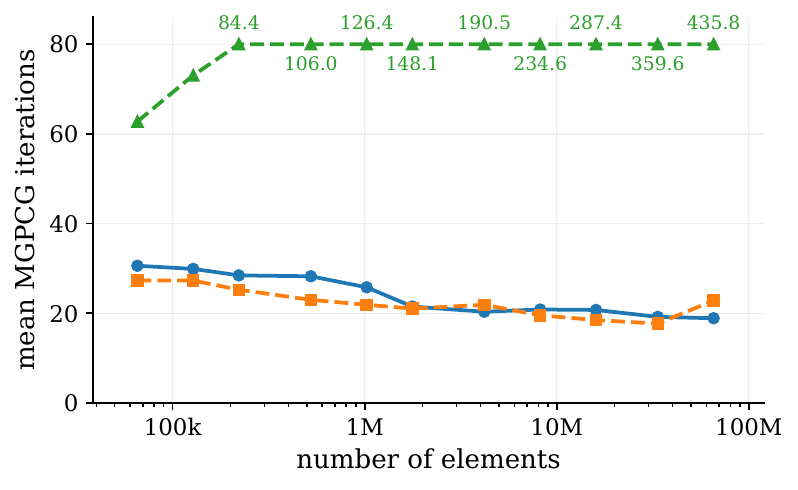}
    \caption{
        Comparison of \topax{}, Futtop, and pyFANTOM.
        Left: runtime for 20 design iterations in seconds; annotations show the
        speedup of \topax{} over the fastest available competing implementation.
        Center: peak device-memory consumption in GiB.
        Right: mean stiffness-MGPCG iteration count over the 20 design iterations;
        values above 80 are clipped and annotated with their actual values.
    }
    \label{fig:traeff_gpu}
\end{figure*}

Figure~\ref{fig:traeff_gpu} compares the runtime for 20 design iterations, the peak device-memory consumption of the GPU, and the mean number of stiffness-MGPCG iterations. 
\topax{} is the fastest implementation at every considered resolution.
Its speedup relative to the fastest competing implementation ranges from $6.1$ to $13.6$.
For the largest problem, containing $65.5$ million elements, \topax{} completes the 20 design iterations in $99.6\,\mathrm{s}$, compared with $720.0\,\mathrm{s}$ for Futtop and $9595.8\,\mathrm{s}$ for pyFANTOM.
This corresponds to speedups of $7.2$ and $96.4$, respectively.

The device-memory results show the same general trend at the larger resolutions.
For the largest mesh, containing $65.5$ million elements, the peak device-memory consumption is $33.5\,\mathrm{GiB}$ for \topax{}, $76.5\,\mathrm{GiB}$ for Futtop, and $74.4\,\mathrm{GiB}$ for pyFANTOM.
Futtop and pyFANTOM therefore require approximately $2.28$ and $2.22$ times as much device memory as \topax{}, respectively.
The memory scaling and its relation to the different representations of the Galerkin hierarchy are examined in more detail below.

The differences in MGPCG convergence are particularly pronounced for pyFANTOM.
Its mean iteration count increases from $62.8$ for the smallest mesh to $435.75$ for the $640\times320\times320$ mesh.
Over the same range, the mean iteration count for \topax{} decreases from $30.6$ to $20.8$.
Futtop remains much closer to \topax{} in this respect.
On the largest mesh, for example, \topax{} requires an average of $18.9$ stiffness-MGPCG iterations per design iteration, compared with $22.85$ for Futtop.
The runtime difference between \topax{} and Futtop therefore cannot be explained by a
systematically smaller number of MGPCG iterations alone.
In contrast, the increasing iteration count in pyFANTOM contributes substantially to its
runtime growth, although its more expensive multigrid cycle also accounts for part of the difference.

\subsubsection*{Details on the representations of the Galerkin hierarchy and the preconditioners}

All three implementations solve the stiffness systems using MGPCG with four grid resolutions and a relative tolerance of $10^{-5}$. Their coarse operators are obtained by Galerkin projection, \cf Equation~\ref{eq:galerkin}.
Although the algebraic construction is common to all three implementations, they differ in how the resulting operators are represented and applied.

In \topax{}, the stiffness operator is represented differently at each level of the hierarchy.
On the finest grid, the stiffness action is evaluated matrix-free using the slicing strategy described in Section~\ref{sec:slicing_matrix_free}.
On the first coarsened grid, the Galerkin operator is applied implicitly using the batched formulation described in Section~\ref{sec:sliced_galerkin}.
On the second coarsened grid, the individual stiffness matrices of the coarse elements are constructed explicitly.
On the coarsest grid, the global stiffness matrix is assembled in CSR format and solved by a direct cuDSS factorization.
One weighted-Jacobi step with a damping parameter of $0.6$ is applied before and after each coarse-grid correction.

Both \topax{} and Futtop apply the fine-grid stiffness operator without assembling a global stiffness matrix.
Futtop implements this operation using explicit node and element indexing: for each node, it identifies the adjacent elements, gathers their local displacement degrees of freedom, and sums the corresponding stiffness contributions.
In contrast, \topax{} expresses the same operation using convolutional primitives and spatial slicing, without explicitly constructing or traversing element-to-node indices.
The two approaches also differ in their treatment of the first coarsened
grid.
\topax{} retains this grid as an active level of the multigrid cycle and applies its Galerkin operator implicitly.
Futtop does not evaluate a stiffness operator or perform smoothing at this level; instead, it restricts the residual directly to the next coarser grid.
The first coarsened grid in Futtop therefore serves only as an intermediate stage for the transfer operators.
On the second coarsened level, Futtop, like \topax{}, uses an explicit Galerkin operator obtained by materializing the coarse element matrices.
This operator is coarsened once more to form the coarsest system.
Unlike \topax{}, which solves this system directly, Futtop applies Jacobi-preconditioned conjugate gradients with a relative tolerance of $10^{-10}$ and a maximum of 800 iterations.
For smoothing, Futtop uses nodal symmetric successive over-relaxation,
adapted from \citet{wu2015system}, rather than the weighted-Jacobi scheme
used by \topax{}.
Note that Futtop uses mixed precision, storing its auxiliary MGPCG vectors in single
precision while evaluating the stiffness operators and multigrid hierarchy
in double precision. Both \topax{} and pyFANTOM use double precision
throughout.

Similar to Futtop, pyFANTOM constructs element-to-node connectivity and node-to-element lookup arrays, which it uses to evaluate the fine-grid stiffness action.
As in \topax{}, pyFANTOM retains the first coarsened grid as an active part of the multigrid cycle.
Unlike \topax{}, however, pyFANTOM explicitly constructs and stores the Galerkin operator at this level in CSR format.
The operators on all subsequent coarse grids are likewise constructed and stored explicitly.

During each application of the multigrid preconditioner, pyFANTOM computes the correction from the two deepest grids twice, resulting in a W-cycle rather than the V-cycle used by \topax{}.
On the coarsest grid, the assembled system is transferred to the CPU and solved by a direct CHOLMOD factorization.
For smoothing, pyFANTOM applies three weighted-Jacobi steps before and after each coarse-grid correction, with the damping parameter adjusted adaptively.

These different representations are reflected in the device-memory scaling.
At the smallest resolutions, pyFANTOM uses less device memory than \topax{}, for which the fixed memory associated with the runtime and compiled program is comparatively prominent.
The curves cross between the $96\times48\times48$ and $128\times64\times64$ meshes.
At $400\times200\times200$, pyFANTOM uses $18.9\,\mathrm{GiB}$, compared with $8.2\,\mathrm{GiB}$ for \topax{}, a factor of approximately $2.29$.
The growing separation is consistent with pyFANTOM explicitly storing the Galerkin operator on every coarse grid, including the first coarsened one.

The three implementations thus adopt distinct treatments of the first coarsened grid.
pyFANTOM retains this grid as an active level but explicitly stores the corresponding Galerkin operator in sparse format.
Futtop avoids this storage by using the grid only as an intermediate stage for transferring the residual to the next coarser level, without applying smoothing or an independent coarse-grid correction.
Like pyFANTOM, \topax{} also retains the first coarsened grid as an active level, but evaluates its Galerkin operator implicitly.
In the \topax{} implementation of the benchmark, coarse-element stiffness matrices are materialized only from the second coarsened grid onward, and the global sparse matrix required by the direct solver is assembled only on the coarsest grid.
The measurements indicate that this strategy avoids the memory growth associated with an explicitly stored operator on the first coarsened grid while preserving its role in the multigrid cycle.

\subsubsection*{Computational setup and measurements}

All benchmark runs used one NVIDIA GH200 GPU with approximately $96\,\mathrm{GiB}$ of device memory and were allocated 72 Grace CPU cores.
The three implementations were run on the same type of compute node, but on separate nodes and in separate software environments.

The \topax{} measurements used Python~3.12.14 and JAX~0.11.1 with the CUDA backend. Futtop was built from the released Futhark implementation and used the OpenCL backend. The pyFANTOM measurements used version~0.1.0 with Python~3.9.25 and CuPy~13.6.0.
Its CPU coarse-grid solver used CHOLMOD with METIS~\citep{karypis1998fast} support.
The software environments were configured independently according to the dependencies of each implementation.

The reported runtimes include only the 20 design iterations.
For \topax{}, setup, jit compilation, and warm-up are excluded.
For Futtop and pyFANTOM, the runtime is obtained by summing the 20 reported design-iteration times, excluding initialization preceding the first iteration.
All work performed within a design iteration remains included.
A single measurement was used for each completed mesh and implementation; the reported values are therefore individual observations rather than averages over repeated runs.

Device-memory usage was sampled at intervals of $0.2\,\mathrm{s}$.
The reported value is the maximum sampled consumption after subtracting the baseline measured immediately before the corresponding run.
The \topax{} memory measurements were performed separately over ten design iterations with JAX device-memory preallocation disabled.
The memory results include device memory only and exclude host allocations.
This distinction is particularly relevant for pyFANTOM because its coarsest matrix and CHOLMOD factorization reside in host memory.
Even when excluding this host-side memory, pyFANTOM requires approximately $2.22$ times as much device memory for the largest problem.

The measured values describe the device memory reserved by each process rather than an analytical sum of simultaneously live arrays.
In particular, the Futhark OpenCL runtime retains device allocations for reuse.
The Futtop measurements therefore include cached allocations that remain reserved by the process even when they are no longer actively used by the algorithm.
Although this may produce a larger value than an analytical accounting of the working set, it represents the amount of device memory occupied during execution.

\subsubsection*{Multiresolution approximation of the Galerkin hierarchy}

To reduce the memory required to represent exact Galerkin coarse-grid operators, \citet{traff2023simple} also propose a multiresolution approximation.
Instead of constructing the coarse-element matrices through Galerkin projection, the stiffness matrix of each coarse element is evaluated from the spatially varying fine-grid densities contained within it, using the multiresolution element formulation of \citet{nguyen2010computational}.
The resulting operators therefore retain information from the fine-grid material distribution but are not exact Galerkin projections, giving rise to a slightly worse preconditioner~\citep{traff2023simple}.
In the provided implementation, the two deepest grids are stored and processed on the CPU.
On the second coarsened grid, the materialized coarse-element matrices are stored and used to evaluate the stiffness action without assembling a global sparse matrix.
On the coarsest grid, the global stiffness matrix is assembled and the resulting system is solved by a direct CHOLMOD factorization.
This shifts a substantial part of the memory required by the multigrid hierarchy from the GPU to the host.

For the largest problem, the multiresolution preconditioner, which is implemented with mixed precision, used $13.68\,\mathrm{GiB}$ of GPU memory and $23.76\,\mathrm{GiB}$ of CPU memory during steady-state operation.
This gives a combined footprint of $37.45\,\mathrm{GiB}$.
The corresponding values for \topax{} are $31.60\,\mathrm{GiB}$ (GPU) and $5.35\,\mathrm{GiB}$ (CPU), giving $36.96\,\mathrm{GiB}$ in total.
The two approaches therefore have nearly identical combined memory requirements, although the memory is distributed very differently between the CPU and GPU.
These measurements show that comparable memory efficiency does not require approximating the Galerkin hierarchy, but can instead be achieved through implicit operator application, delayed materialization, and careful management of initialization data.

These memory measurements were obtained using the released OpenMP~\citep{dagum1998openmp} implementation of the multiresolution formulation\footnote{\href{https://github.com/topopt/TopOpt-in-OpenMP-GPU}{https://github.com/topopt/TopOpt-in-OpenMP-GPU}}. 
Although the implementation could be executed on the GH200, only the two smallest problems completed all 20 design iterations.
For the $128\times64\times64$ problem, the MGPCG iteration count increased from 55 at design iteration~8 to 105 at iteration~9, and the solver reached its hard-coded limit of 1000 iterations at design iterations~10 and~11.
We investigated this loss of convergence by recording $\mathbf{r}^{\mathsf{T}}\mathbf{z}$ after each application of the multigrid V-cycle.
Here, $\mathbf{r}$ is the current conjugate gradient residual and $\mathbf z=\mathcal G(\mathbf r,0)$ is the corresponding preconditioned residual in Algorithm~\ref{alg:mgcg} (Lines~\ref{alg:mgcg:line:precon_residual_num} and~\ref{alg:mgcg:line:precon_residual_denom}).
For a nonzero residual, positive definiteness requires $\mathbf{r}^{\mathsf{T}}\mathbf{z}>0$.
However, in the seventh MGPCG iteration of design iteration~8, we obtained $\mathbf{r}^{\mathsf{T}}\mathbf{z}=-3.42\times10^{-4}$.
The multigrid cycle therefore did not define a positive-definite preconditioner at this point.
We consequently exclude this implementation from the runtime comparison.

The memory measurements are not affected by the failure of the iterative solver.
The principal arrays, coarse-grid operators, and CHOLMOD factorization have mesh-dependent dimensions and were allocated before the preconditioner deteriorated.
Their host and device footprints reached stable values during the measurement interval. Moreover, the two smallest problems completed normally, and the measured GPU-memory consumption agrees closely with that reported by \citet{traff2023simple}.
We therefore retain the memory results as representative of the implementation, despite not reporting its runtime.

\subsubsection*{Trading off compute and memory}

The preceding comparison used a hybrid representation in which the first coarsened operator is applied implicitly and the operators on deeper grids are materialized. This choice is not fixed by the method.
Applying an operator implicitly avoids storing a separate stiffness matrix for every coarse element, but its contributions must be reconstructed whenever the operator is used.
Materialization incurs additional storage during setup and throughout the optimization, while reducing the work required during each multigrid cycle.

To quantify this trade-off, we compare two \topax{} configurations on the benchmark introduced above. Both use four grid resolutions and differ only in their treatment of the second coarsened grid.
We denote these configurations by L4-C2 and L4-C3, respectively.
The L4 part expresses that there are four levels in total (including the fine grid at $\ell=0$), and C2/C3 denotes the first level at which the Galerkin operator is materialized.
This means that in the L4-C3 configuration, both intermediate Galerkin operators are applied implicitly, and materialization is delayed until the coarsest operator is formed for the direct solve.
The L4-C2 configuration is already used in the preceding comparison with the first Galerkin operator being implicit and the second being materialized.

\begin{figure*}[h]
    \centering
    \includegraphics[width=0.32\textwidth]
        {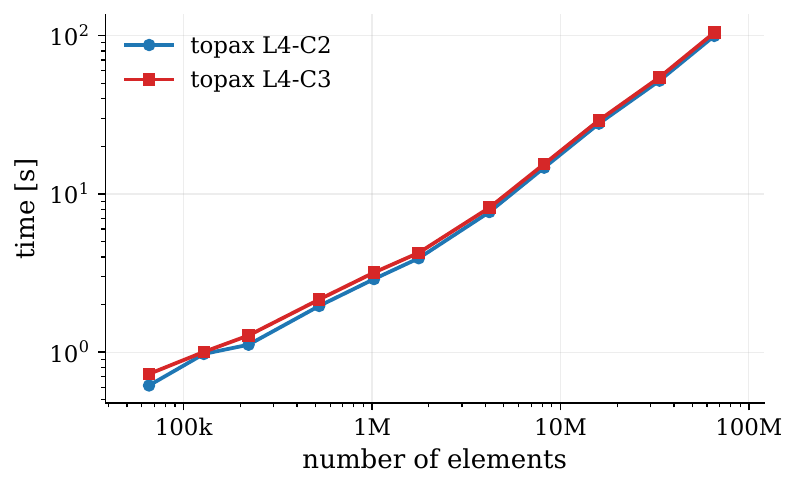}
    \hfill
    \includegraphics[width=0.32\textwidth]
        {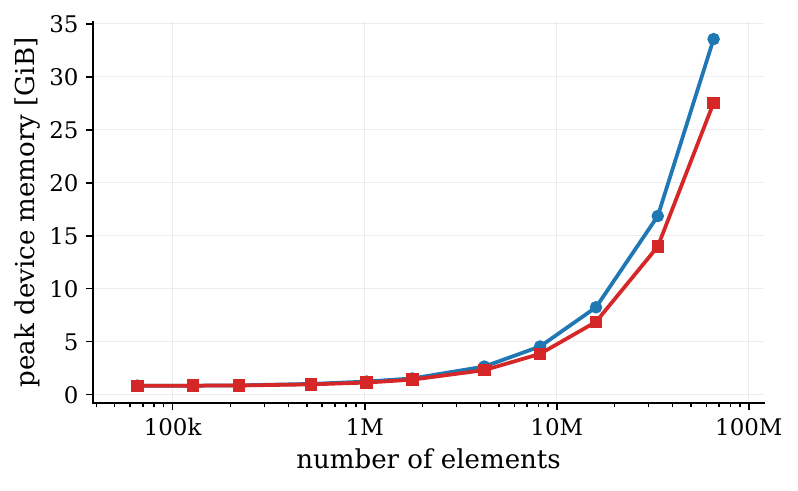}
    \hfill
    \includegraphics[width=0.32\textwidth]
        {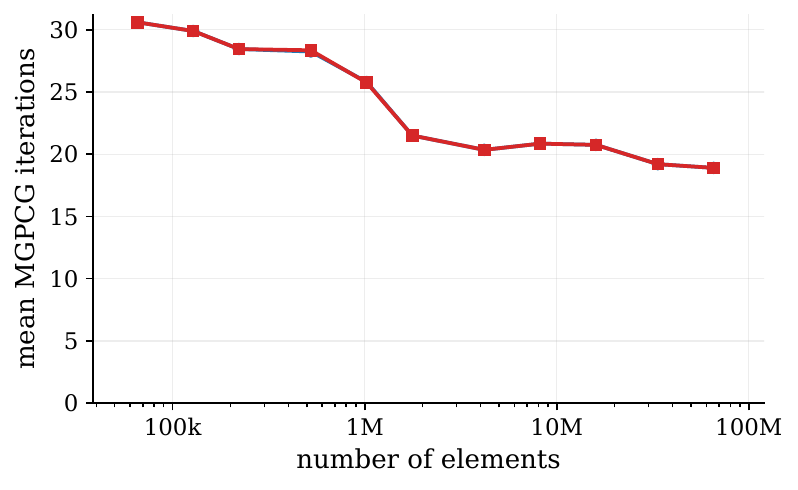}
    \caption{
        Comparison of two \topax{} Galerkin hierarchies for the \citet{traff2023simple} benchmark.
        Both configurations use four grid resolutions; L4-C2 materializes the coarse-element matrices after the second coarsening, whereas L4-C3 delays materialization until the third and final coarsening.
        Left: runtime for 20 design iterations.
        Middle: peak device-memory consumption.
        Right: mean stiffness-MGPCG iteration count over the 20 design iterations.
    }
    \label{fig:traeff_gpu_topax}
\end{figure*}

Figure~\ref{fig:traeff_gpu_topax} shows the runtime, peak device-memory consumption, and mean stiffness-MGPCG iteration count for both configurations.
Their MGPCG iteration counts are effectively identical because they represent the same four-level Galerkin hierarchy and differ only in how one of its operators is evaluated.
The observed differences in runtime and memory can therefore be attributed to the operator representation rather than to changes in preconditioner quality.

Materializing the second coarsened operator gives the shorter runtime at every tested resolution.
For the largest problem, L4-C2 completes the 20 design iterations in $99.6\,\mathrm{s}$, compared with $104.4\,\mathrm{s}$ for L4-C3. 
Delaying materialization therefore increases the runtime by approximately $4.9\%$ at this resolution.
However, implicit representation avoids storing spatially varying coarse-element matrices, reducing the maximum device-memory consumption from $33.5\,\mathrm{GiB}$ to $27.5\,\mathrm{GiB}$.
This corresponds to a reduction of approximately $6.0\,\mathrm{GiB}$, or $17.9\%$.

The comparison demonstrates that the crossover level can be selected according to the available device memory.
Materializing the operator earlier is preferable when sufficient memory is available, since it avoids repeatedly evaluating the implicit Galerkin contraction.
Delaying materialization provides a lower-memory alternative without changing the multigrid hierarchy or degrading MGPCG convergence.
For the largest problem considered here, this reduces peak device-memory consumption by almost one fifth at a runtime cost of less than five percent.

\subsection{Memory-conscious multigrid for large-scale topology optimization}
\label{sec:memory_conscious_large_scale}

The preceding experiments show that the memory occupied by a Galerkin hierarchy depends strongly on how its coarse operators are represented.
This becomes particularly important for three-dimensional problems with hundreds of millions of elements, where the combined memory required by the multigrid hierarchy, the fine-grid fields, and the temporary arrays used during operator evaluation may lead to exhaustion of the available memory.

We therefore compare \topax{} with TOP3D\_XL, the matrix-free MATLAB
implementation of \citet{wang2025efficient}.
TOP3D\_XL was designed specifically to limit memory consumption at this scale.
In addition to avoiding assembly of the fine-grid stiffness matrix, it reduces the size of the Galerkin hierarchy by replacing the first two conventional factor-two coarsenings with a single factor-four transfer.
This removes the first grid that would appear in a dyadically coarsened hierarchy.
In contrast, \topax{} retains this grid as an active correction level but applies its Galerkin operator implicitly.
The comparison therefore examines whether the memory reduction obtained by omitting a grid can instead be achieved through a memory-efficient representation of its operator.

\subsubsection*{The benchmark problem}

The benchmark is the rectangular cuboid cantilever considered by
\citet{wang2025efficient}.
The design domain has fixed physical dimensions $1\times0.5\times0.5$.
All displacement components are constrained on the $x=0$ face.
A distributed load with a total magnitude of one acts in the negative $z$-direction on the opposite face.
The loading and boundary conditions follow the original benchmark; their complete definitions are available in the cited paper and in the source code accompanying this work.

We consider meshes containing $96\times48\times48$ (R96), $192\times96\times96$ (R192), $384\times192\times192$ (R384), and $800\times400\times400$ (R800) elements.
The largest mesh contains $128$ million elements and approximately $386.4$ million displacement degrees of freedom.
The prescribed volume fraction is $0.12$.
The SIMP and material parameters are the same as in Section~\ref{sec:large_scale_galerkin_gpu}: $p=3$, $E=1$, $\nu=0.3$, and a minimum relative stiffness of $10^{-6}$.

Following the classical sensitivity-filter formulation of \citet{sigmund200199}, implemented through the Helmholtz-type PDE filter~\citep{lazarov2011filters}, the unfiltered design densities enter the stiffness interpolation, while the density-weighted compliance sensitivities are regularized before the optimality-criteria update.
Mesh dependence is controlled by using a filter radius of $\sqrt{3}$ element widths.
The design is updated using the optimality-criteria method with a move limit of $0.2$, and each problem is run for 50 design iterations.
Both implementations use a relative residual tolerance of $10^{-3}$ for the stiffness solve and $10^{-4}$ for the Helmholtz-filter solve.
The released TOP3D\_XL implementation permits at most 800 MGPCG iterations, whereas the corresponding limit in \topax{} is 600; neither limit is reached in the completed measurements.

\subsubsection*{CPU evaluation}

TOP3D\_XL evaluates the fine-grid stiffness action without assembling a global stiffness matrix.
Its adapted multigrid cycle reduces the size of the hierarchy by omitting the grid produced by the first conventional factor-two coarsening.
Instead, the transfer operator maps directly between the fine grid and a grid with four times the element width in each spatial direction.
Each element on the resulting coarse grid therefore contains $4^3=64$ fine elements, and its stiffness matrix is obtained by Galerkin projection using the corresponding factor-four interpolation operator.
Subsequent grids are obtained by conventional factor-two coarsening.
This resembles the treatment of the first coarsened grid in Futtop, but TOP3D\_XL constructs the transfer directly between the fine grid and the factor-four grid rather than passing through an intermediate factor-two grid.

TOP3D\_XL selects the number of grid resolutions adaptively according to the problem size.
For R96, R192, and R384, it uses three grid levels and therefore performs two coarsenings.
For R800, it adds another coarse grid and uses four grid resolutions.
The resulting R800 hierarchy is
\begin{align*}
    800\times400\times400
    \;\longrightarrow\;
    200\times100\times100
    \;\longrightarrow\;
    100\times50\times50
    \;\longrightarrow\;
    50\times25\times25.
\end{align*}
On the intermediate grids the element-wise stiffness matrices are materialized, while on the coarsest grid, the global stiffness matrix is assembled and factorized using CHOLMOD.

For \topax{}, we consider the following two dyadic hierarchies for R800:
\begin{align*}
    800\times400\times400
    &\;\longrightarrow\;
    400\times200\times200
    \;\longrightarrow\;
    200\times100\times100
    \;\longrightarrow\;
    100\times50\times50,
    \\
    800\times400\times400
    &\;\longrightarrow\;
    400\times200\times200
    \;\longrightarrow\;
    200\times100\times100
    \;\longrightarrow\;
    100\times50\times50
    \;\longrightarrow\;
    50\times25\times25.
\end{align*}
We denote these configurations by L4-C3 and L5-C4, respectively.
In both cases, all intermediate Galerkin operators are applied implicitly, while the global matrix is assembled only on the coarsest level for the CHOLMOD solve.
For R800, L4-C3 contains the same number of levels as TOP3D\_XL, but its coarsest grid has twice as many elements along each spatial direction.
L5-C4 adds one level and reaches the same coarsest-grid resolution as TOP3D\_XL.

\begin{figure*}[h]
    \centering
    \includegraphics[width=0.32\textwidth]
        {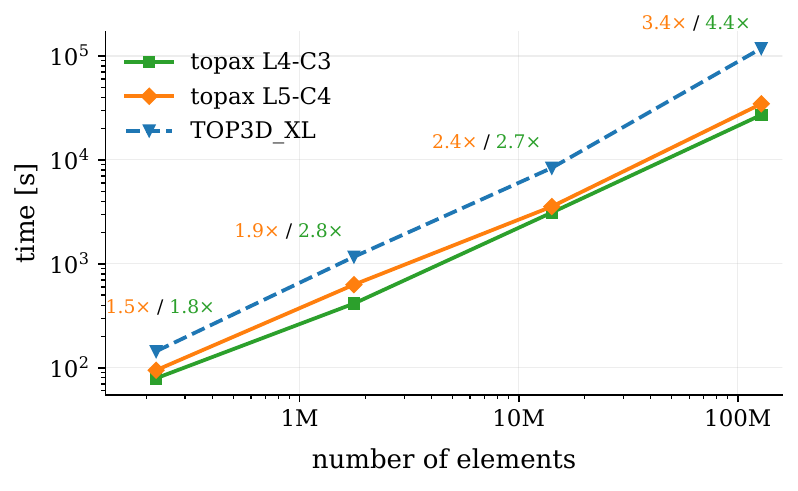}
    \hfill
    \includegraphics[width=0.32\textwidth]
        {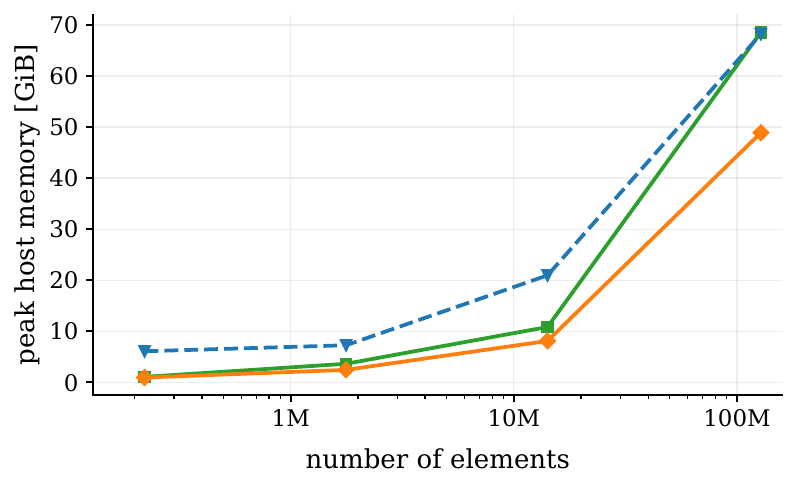}
    \hfill
    \includegraphics[width=0.32\textwidth]
        {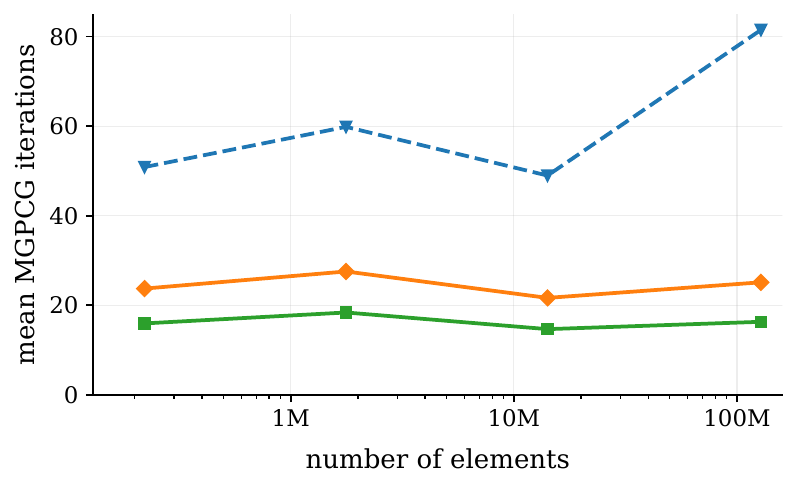}
    \caption{
        CPU comparison of TOP3D\_XL and the L4-C3 and L5-C4 \topax{} hierarchies.
        Left: runtime for 50 design iterations; annotations give the speedups of L5-C4 and L4-C3, respectively, over TOP3D\_XL.
        Center: peak CPU memory.
        Right: mean stiffness-MGPCG iteration count over the 50 design iterations.
    }
    \label{fig:wang_cpu}
\end{figure*}

Figure~\ref{fig:wang_cpu} compares the runtime, peak CPU memory, and mean stiffness-MGPCG iteration count.
All measurements were performed on CPU nodes containing two 128-core AMD EPYC 9755 processors operating at $2.7\,\mathrm{GHz}$ and $768\,\mathrm{GB}$ of DDR5 memory.
Simultaneous multithreading was disabled.
Each benchmark run was allocated 32 physical CPU cores and approximately $86.4\,\mathrm{GiB}$ of host memory, which replicates the setup by \citet{wang2025efficient}.
The implementations were run on the same node type, but individual measurements were obtained on separate nodes.
TOP3D\_XL was executed using MATLAB R2025b, with the OpenMP and MKL thread counts set to 32.
The \topax{} measurements used Python~3.12.14 and JAX~0.11.1 with the CPU backend.

The L4-C3 hierarchy is faster than L5-C4, whereas the additional coarsening in L5-C4 reduces memory consumption.
Across the displayed resolutions, the speedup over TOP3D\_XL ranges from $1.8$ to $4.4$ for L4-C3 and from $1.5$ to $3.4$ for L5-C4.
For the largest problem, containing $128$ million elements, TOP3D\_XL requires $118\,149.0\,\mathrm{s}$ ($32.8\,\mathrm{hrs}$) for the 50 design iterations.
The corresponding \topax{} runtimes are $27\,058.8\,\mathrm{s}$ ($7.8\,\mathrm{hrs}$) for L4-C3 and $34\,561.8\,\mathrm{s}$ ($9.6\,\mathrm{hrs}$) for L5-C4, giving speedups of $4.4$ and $3.4$, respectively.
This difference is accompanied by substantially different convergence of the stiffness solver.
TOP3D\_XL requires an average of $81.48$ MGPCG iterations per design iteration, compared with $16.30$ for L4-C3 and $25.12$ for L5-C4.
Retaining the intermediate factor-two grids as active correction levels therefore produces a considerably stronger preconditioner than the non-dyadic hierarchy used by TOP3D\_XL.

For R800, the peak CPU memory sits at $68.20\,\mathrm{GiB}$ for TOP3D\_XL, $68.56\,\mathrm{GiB}$ for L4-C3, and $48.88\,\mathrm{GiB}$ for L5-C4.
Thus, L4-C3 has essentially the same memory requirement as TOP3D\_XL despite factorizing a coarsest system containing eight times as many elements, while L5-C4 requires approximately $28\%$ less host memory.
At every smaller displayed resolution, both \topax{} configurations also require less memory than TOP3D\_XL.
The closing of the gap at R800 coincides with a change in the adaptive TOP3D\_XL hierarchy.
For R96, R192, and R384, TOP3D\_XL uses three grid resolutions and reaches the same coarsest resolution as L4-C3.
At R800, it introduces a fourth grid resolution, moving its direct solve to a $50\times25\times25$ grid, whereas L4-C3 terminates at $100\times50\times50$.
The resulting reduction in the size of the directly factorized system largely explains why the memory consumption of TOP3D\_XL approaches that of L4-C3.
L5-C4 also terminates at $50\times25\times25$, but retains all intermediate factor-two corrections while using substantially less memory than TOP3D\_XL.

These results show that omitting the first factor-two grid is not necessary to keep an exact Galerkin hierarchy tractable.
Applying the intermediate Galerkin operators implicitly retains their contribution to the multigrid cycle without explicitly storing them.
The two \topax{} configurations illustrate the resulting trade-off: the larger directly factorized coarse system in L4-C3 improves convergence and runtime, whereas the additional implicit level in L5-C4 substantially reduces the memory requirement.

The reported runtimes include only the 50 design iterations.
For \topax{}, initialization, JIT compilation, and warm-up are excluded.
The TOP3D\_XL runtime is obtained by summing its reported design-iteration times.
Peak memory is the maximum resident-set size of the complete process.
A single completed measurement is used for each mesh and implementation.

\citet{wang2025efficient} report that the $128$-million-element problem can be executed on a computer with $64\,\mathrm{GB}$ of RAM.
We could not reproduce this bound using the released implementation: the corresponding run reached a peak resident-set size of $68.20\,\mathrm{GiB}$.
Our measurement was performed with a larger memory allocation and did not involve swapping.
The difference may arise from the MATLAB version, the software environment, or a different accounting of memory consumption.

\subsubsection*{GPU scaling}

We additionally evaluate the L4-C3 and L5-C4 configurations using the GPU backend of \topax{}.
The numerical formulation and multigrid hierarchies are identical to those used for the CPU measurements; only the execution backend and the coarsest direct solver differ.
On the GPU, the coarsest matrix is factorized using cuDSS instead of CHOLMOD.
The measurements use the same NVIDIA GH200 system and the same device-memory measurement procedure as in Section~\ref{sec:large_scale_galerkin_gpu}.
Runtime measurements contain all 50 design iterations, while device-memory consumption is measured separately over ten design iterations with JAX memory preallocation disabled.

\begin{figure*}[h]
    \centering
    \includegraphics[width=0.32\textwidth]
        {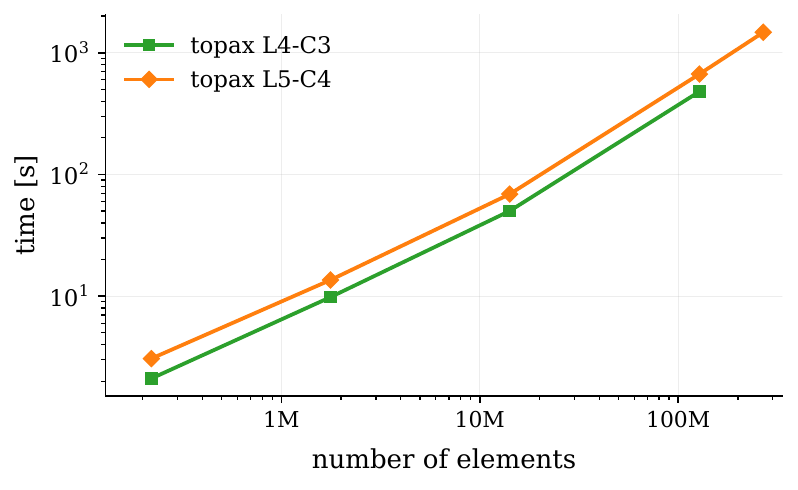}
    \hfill
    \includegraphics[width=0.32\textwidth]
        {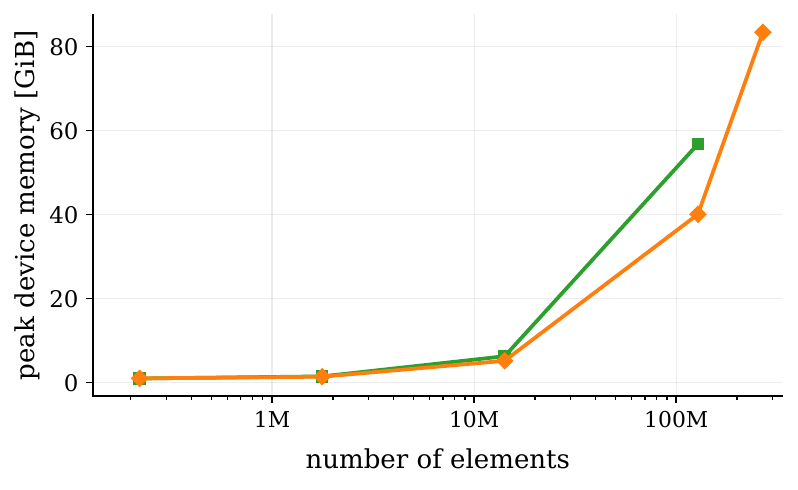}
    \hfill
    \includegraphics[width=0.32\textwidth]
        {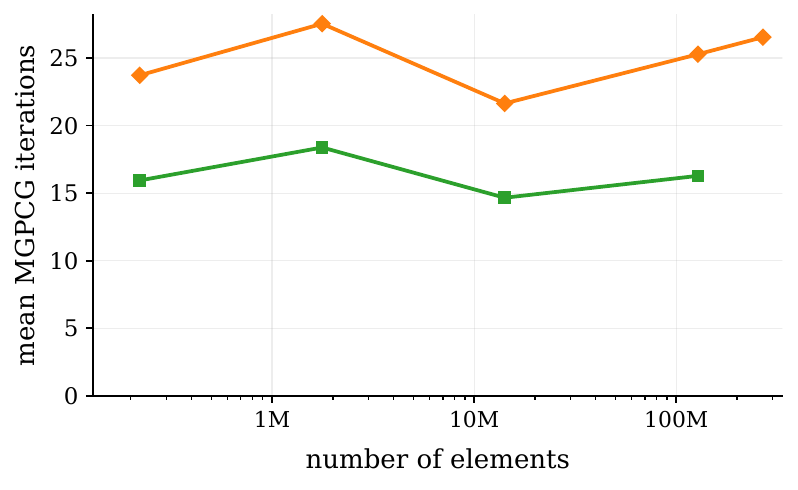}
    \caption{
        GPU scaling of the L4-C3 and L5-C4 \topax{} hierarchies.
        Left: runtime for 50 design iterations.
        Center: peak GPU memory consumption.
        Right: mean stiffness-MGPCG iteration count over the 50 design iterations.
    }
    \label{fig:wang_gpu}
\end{figure*}

Figure~\ref{fig:wang_gpu} compares the runtime,  peak GPU memory consumption, and mean stiffness-MGPCG iteration count.
At the resolutions available for both configurations, L4-C3 is consistently faster and requires fewer MGPCG iterations.
L5-C4, however, substantially reduces memory consumption at the larger resolutions by moving the direct solve to a smaller coarsest grid.

For R800, L4-C3 completes the 50 design iterations in $479.8\,\mathrm{s}$, compared with $667.7\,\mathrm{s}$ for L5-C4.
The corresponding mean MGPCG iteration counts are $16.28$ and $25.28$.
L5-C4 is therefore approximately $39\%$ slower at this resolution, but reduces peak device-memory consumption from $56.79\,\mathrm{GiB}$ to $40.07\,\mathrm{GiB}$, a reduction of approximately $29\%$.
The lower memory requirement of L5-C4 enables the R1024 ($1024\times512\times512$) problem, containing $268.4$ million elements and approximately $809$ million displacement degrees of freedom, to fit on a single GH200.
This problem completes 50 design iterations in $1\,471.3\,\mathrm{s}$, or approximately $24.5\,\mathrm{min}$, with a mean of $26.54$ MGPCG iterations per design iteration.
Its peak GPU memory consumption is $83.39\,\mathrm{GiB}$.
The number of elements is $2.10$ times that of R800, while the measured memory consumption is $2.08$ times larger, indicating approximately linear memory scaling.

Across problem sizes, the mean MGPCG iteration count remains between $14.66$ and $18.38$ for L4-C3 and between $21.64$ and $27.54$ for L5-C4.
The iteration counts therefore remain approximately independent of the mesh resolution.
The comparison exposes the same trade-off observed on the CPU: L4-C3 provides the stronger preconditioner and the shorter runtime, whereas the additional implicit level in L5-C4 reduces the size of the directly factorized system and permits substantially larger problems within the available device memory.

\subsection{Large-scale three-dimensional designs}

\begin{figure*}[h]
    \centering
    \includegraphics[width=0.49\textwidth]
        {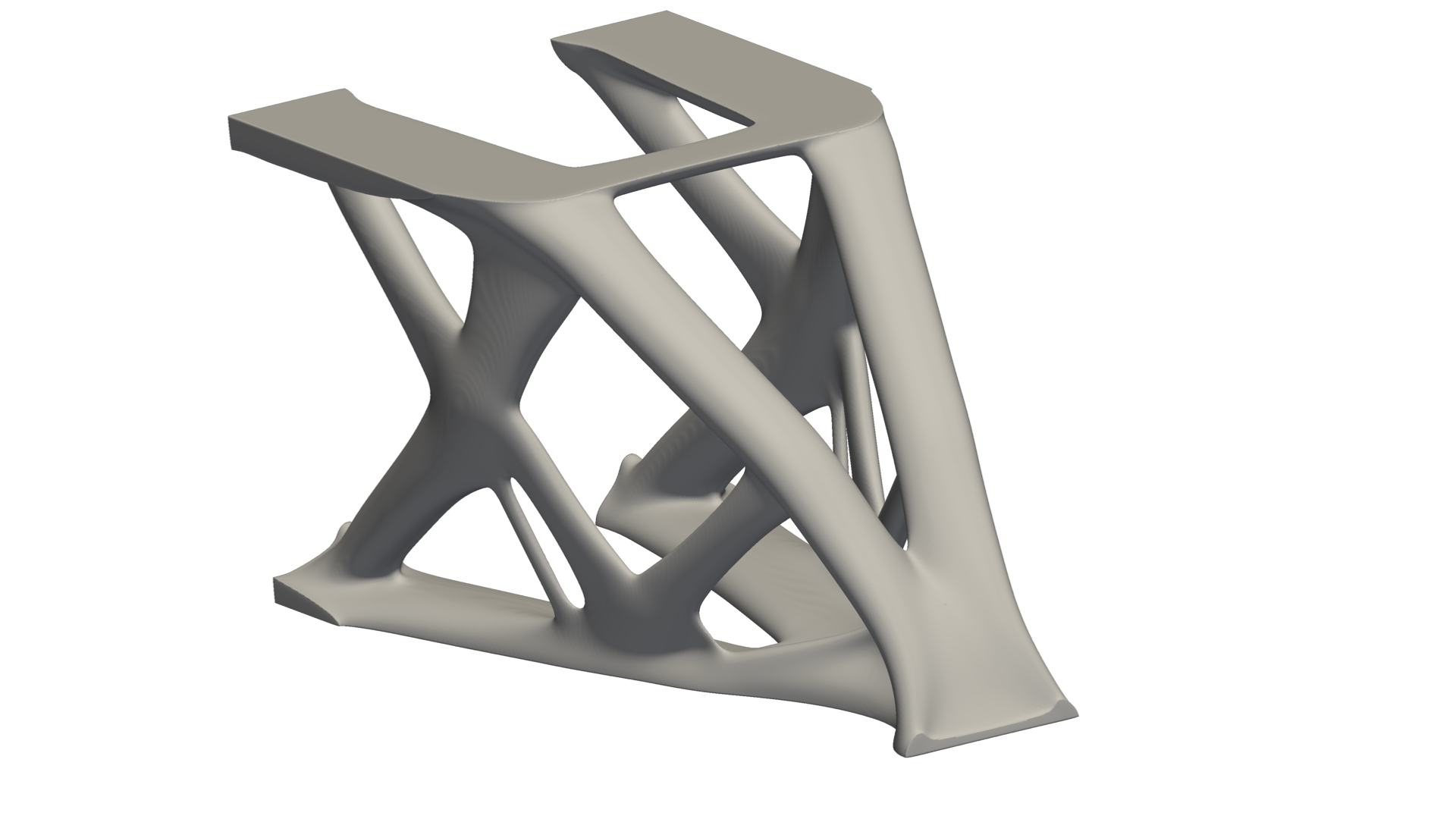}
    \hfill
    \includegraphics[width=0.49\textwidth]
        {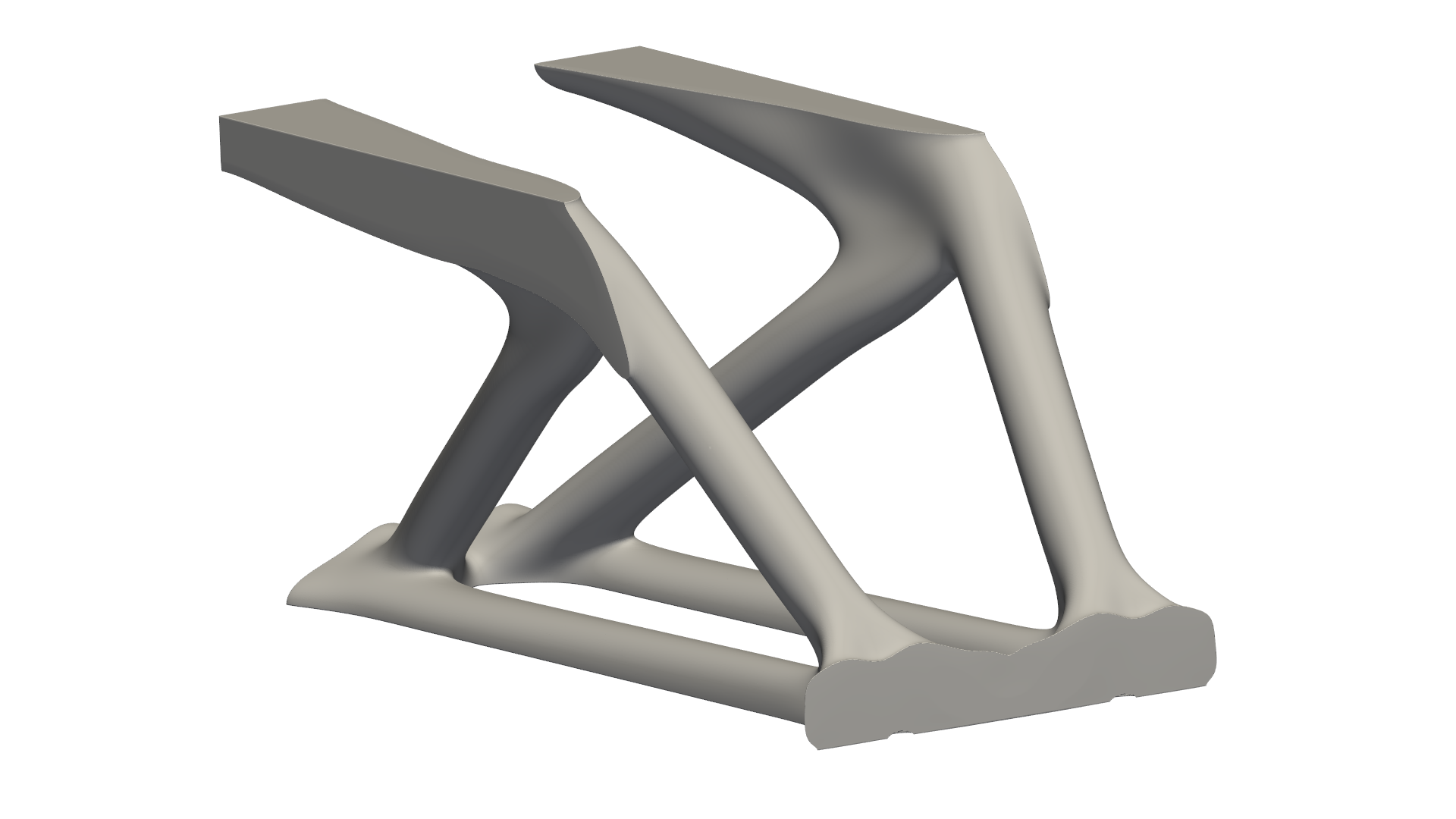}
    \caption{
        Large-scale three-dimensional cantilever designs after 50 design
        iterations.
        Left: the cantilever studied by \citet{traff2023simple}, discretized using $640\times320\times320$ elements with a prescribed volume fraction of $0.1$.
        Right: the cantilever studied by \citet{wang2025efficient}, discretized using $800\times400\times400$ elements with a prescribed volume fraction of $0.12$.
    }
    \label{fig:final_designs}
\end{figure*}

Figure~\ref{fig:final_designs} shows optimized designs obtained by repeating the two cantilever problems introduced in Sections~\ref{sec:large_scale_galerkin_gpu} and \ref{sec:memory_conscious_large_scale} at resolutions of $640\times320\times320$ and $800\times400\times400$, respectively, using different filter radii from those employed in the performance benchmarks.
In the performance benchmarks, the filter radius is kept fixed in element units. For the designs shown here, the radius is instead adjusted so that its physical size matches that used for the corresponding benchmark on the coarsest mesh.
For the cantilever studied in Section~\ref{sec:large_scale_galerkin_gpu}, the radius of $1.5$ element widths at a resolution of $64\times32\times32$ becomes $15$ element widths at $640\times320\times320$.
For the cantilever studied in Section~\ref{sec:memory_conscious_large_scale}, the Helmholtz-type PDE filter radius of $\sqrt{3}$ element widths at $48\times24\times24$ becomes $17\sqrt{3}$ element widths at $800\times400\times400$.

The first design contains $65.5$ million elements and $198.1$ million displacement degrees of freedom, while the second contains $128$ million elements and $386.4$ million displacement degrees of freedom.
Both were optimized for 50 design iterations on a single NVIDIA GH200.
The complete calculations, including setup, compilation, warm-up, and file handling, required approximately four and five minutes, respectively.
These times are lower than those observed in the benchmark runs because the larger filter radii suppress small-scale features and produce smoother density distributions, substantially reducing the number of MGPCG iterations.

The surfaces were extracted from the continuous density fields at $\rho=0.5$ using the Flying Edges algorithm~\citep{schroeder2015flying} as implemented in VTK~\citep{vtkBook} and accessed through PyVista~\citep{sullivan2019pyvista}.
No binarization, connectivity filtering, or geometric smoothing was applied.
These examples demonstrate that three-dimensional topology optimization with hundreds of millions of displacement degrees of freedom can be completed on a single GPU with near-real-time turnaround.

\section{Conclusions}

This work introduces memory-bounded tensor formulations for matrix-free finite-element analysis and Galerkin multigrid on structured Cartesian meshes.
The fine-grid stiffness action is expressed through convolutional primitives and evaluated in overlapping spatial slices, thereby limiting the intermediate memory without altering the underlying finite-element operator.
For the coarse grids, the Galerkin operators are represented through fine-grid stiffness interpolation coefficients and precomputed local operators, allowing their action to be evaluated implicitly without materializing a separate stiffness matrix for every coarse element.
Combining implicit evaluation with delayed materialization provides a configurable trade-off between computational cost and memory consumption while preserving an exact Galerkin hierarchy.
These methods are implemented in \topax{}, a modular JAX-based library for density-based topology optimization that combines hardware-accelerated tensor operations, just-in-time compilation, and automatic differentiation.

The numerical experiments show that these representations translate directly into substantial improvements in runtime and memory consumption.
In the GPU comparison, \topax{} is between $6.1$ and $13.6$ times faster than the fastest competing implementation and requires less than half the device memory for the largest problem.
On the CPU, \topax{} is up to $4.4$ times faster than TOP3D\_XL, while also reducing memory requirements.
The largest experiment contains more than $809$ million displacement degrees of freedom and completes 50 design iterations on a single GPU in approximately $24.5$ minutes, with a peak device-memory consumption of $83.4\,\mathrm{GiB}$.

The present implementation focuses on linear elasticity on uniform Cartesian meshes, but the resulting solver provides a foundation for broader differentiable simulation and optimization workflows.
One immediate direction is to expose the multigrid implementation as a reusable matrix-free linear solver or preconditioner for other PDE-constrained problems, including the linear systems arising within nonlinear solvers.
The largest experiment also approaches the capacity of a single $96\,\mathrm{GiB}$ GPU.
At this resolution, one double-precision vector containing all displacement degrees of freedom occupies approximately $6\,\mathrm{GiB}$.
Several such vectors are required simultaneously by MGPCG, in addition to the design, filtering, sensitivity, and optimizer fields.
These persistent full-resolution arrays impose a memory requirement that cannot be removed by further reducing temporary storage, and larger problems will therefore require distributed execution.
Multi-GPU topology optimization has already been demonstrated using distributed algebraic multigrid~\citep{herrero2021multi}, while distributed-memory implementations using CPUs have reached gigavoxel scales~\citep{aage2017giga,baandrup2020closing}.
Because the proposed operators are local, spatially sliced, and expressed in JAX, which supports multi-device array partitioning, such an extension should primarily constitute an engineering task rather than require a reformulation of the method.
To facilitate reproducibility and further development, we release the complete implementation and benchmark scripts alongside this work.

\section*{Acknowledgements}

Pedro Zuidberg Dos Martires was supported by the Wallenberg AI, Autonomous Systems and Software Program (WASP) funded by the Knut and Alice Wallenberg Foundation, and the Swedish Research Council (Vetenskapsrådet) through a Starting Grant (grant number 2025-06142).
Computational resources were provided by the National Academic Infrastructure for Supercomputing in Sweden (NAISS), funded by the Swedish Research Council.

\section*{Declaration of generative AI usage}

During the preparation of this work, the authors used OpenAI Codex to assist with drafting and revising manuscript text and with improving its clarity and readability.
The technical content, analyses, and conclusions were determined by the authors.
All AI-generated suggestions were reviewed, verified, and edited by the authors, who take full responsibility for the content of the published article.
OpenAI Codex was used as a programming assistant for refactoring, code review, debugging, and the preparation of benchmark and evaluation scripts.
All proposed changes were inspected by the authors and validated through automated tests, numerical regression tests, and benchmark reproduction.

\bibliographystyle{plainnat}
\bibliography{references}

\end{document}